\documentclass[twoside]{IEEEtran}
\usepackage{amsthm,amsmath}
\IEEEoverridecommandlockouts
\usepackage{graphicx, epsfig, epsf}
\usepackage[dvips]{color}   %needed for compiling latex commands in figures
\usepackage{mathrsfs}
\usepackage{amssymb}
\usepackage{cite}
\usepackage{stfloats}

\DeclareMathAlphabet{\mathbit}{OT1}{cmr}{bx}{it}

\newtheorem{defn}{Definition}

\newtheorem{thm}{Theorem}

\newtheorem{lemma}{Lemma}
\newtheorem{prop}{Proposition}

\newtheorem{rem}{Remark}

\renewcommand{\P}[1]{{\mathbb{P}}\left(#1\right)}
\renewcommand{\H}[1]{{\mathbb{H}}\left(#1\right)}

\newcommand{\calC}{{\mathcal{C}}}
\newcommand{\calI}{{\mathcal{I}}}
\newcommand{\dmin}{{d_{\mathrm{min}}}}

\newcommand{\J}[1]{J\!\!\left({#1}\right)}

\newcommand{\IAuCode}{I^{\mathcal{C}}_{\mathrm{a},\mathbf{u}}}
\newcommand{\IEuCode}{I^{\mathcal{C}}_{\mathrm{e},\mathbf{u}}}
\newcommand{\IAxCode}{I^{\mathcal{C}}_{\mathrm{a},\mathbf{x}}}
\newcommand{\IExCode}{I^{\mathcal{C}}_{\mathrm{e},\mathbf{x}}}

\newcommand{\IAyCodeMPCC}{I^{\mathcal{C}_{\mathrm{MPCC}}}_{\mathrm{a},\mathbf{x}_\mathrm{MPCC}}}
\newcommand{\IEyCodeMPCC}{I^{\mathcal{C}_{\mathrm{MPCC}}}_{\mathrm{e},\mathbf{x}_\mathrm{MPCC}}}

\newcommand{\IAxCodell}{I^{\mathcal{C}_l}_{\mathrm{a},\mathbf{u}_l}}
\newcommand{\IExCodell}{I^{\mathcal{C}_l}_{\mathrm{e},\mathbf{u}_l}}
\newcommand{\IEyCodell}{I^{\mathcal{C}_l}_{\mathrm{e},\mathbf{x}_l}}

\newcommand{\IAyCodell}{I^{\mathcal{C}_{l}}_{\mathrm{a},\mathbf{x}_l}}

\newcommand{\IExCodeii}{I^{\mathcal{C}_i}_{\mathrm{e},\mathbf{u}_i}}

\newcommand{\IExCodeff}{I^{\mathcal{C}_\mathrm{acc}}_{\mathrm{e},\mathbf{u}_\mathrm{acc}}}
\newcommand{\IAxCodeff}{I^{\mathcal{C}_\mathrm{acc}}_{\mathrm{a},\mathbf{u}_\mathrm{acc}}}
\newcommand{\IAyCodeff}{I^{\mathcal{C}_\mathrm{acc}}_{\mathrm{a},\mathbf{x}_\mathrm{acc}}}

\title{Analysis and Design of Tuned Turbo Codes}

\author{Christian~Koller~\IEEEmembership{Student~Member,~IEEE},
        Alexandre~Graell~i~Amat~\IEEEmembership{Senior~Member,~IEEE},
        J{\"o}rg~Kliewer~\IEEEmembership{Senior~Member,~IEEE},
        Francesca~Vatta~\IEEEmembership{Member,~IEEE},
        Kamil~S.~Zigangirov~\IEEEmembership{Fellow,~IEEE},
        Daniel~J.~Costello,~Jr.~\IEEEmembership{Life~Fellow,~IEEE}
\thanks{Manuscript received October 25, 2010; revised March 02, 2012; accepted
April 03, 2012. This work was partly supported by NSF grants \mbox{CCF08-30651} and 
\mbox{CCF08-30666}, NASA grant NNX09AI66G, the University of Note Dame Center for
Applied Mathematics, and the Swedish Agency for Innovation Systems (VINNOVA) under 
the \mbox{P36604-1} MAGIC project.}
\thanks{This paper was presented in part at the International Symposium on
Information Theory and its Applications, 2008.}
\thanks{C. Koller and D. J. Costello, Jr. are with the Department of Electrical Engineering, 
University of Notre Dame, Notre Dame, IN 46556 USA (e-mail: ckoller@nd.edu; dcostel1@nd.edu).}
\thanks{A. Graell i Amat is with the Department of Signals and Systems, Chalmers
University of Technology, SE-412 96 Gothenburg, Sweden (e-mail:
alexandre.graell@chalmers.se).}
\thanks{J. Kliewer is with the Klipsch School of Electrical and Computer Engineering,
New Mexico State University, Las Cruces, NM 88003-8001 USA (e-mail: jkliewer@
nmsu.edu).}
\thanks{K. Sh. Zigangirov is with the Institute for Problems of Information Transmission,
Moscow, Russia. He is also with the Department of Electrical Engineering,
University of Notre Dame, Notre Dame, IN 46556 USA, and the Department of
Electrical and Information Technology, Lund University, Lund, Sweden (e-mail:
kamil.zigangirov@eit.lth.se).}
\thanks{F. Vatta is with DI3, Universit{\`a} di Trieste, 34127 Trieste, Italy (e-mail:
vatta@units.it).}
\thanks{Copyright \copyright 2012 IEEE. Personal use of this material is permitted.  
However, permission to use this material for any other purposes must be obtained from 
the IEEE by sending a request to {pubs-permissions@ieee.org}.}
}

\begin{document}

\maketitle %\thispagestyle{empty}

\begin{abstract}
  It has been widely observed that there exists a fundamental
  trade-off between the minimum (Hamming) distance properties and the iterative decoding
  convergence behavior of turbo-like codes. While capacity achieving code
  ensembles typically are asymptotically bad in the sense that their
  minimum distance does not grow linearly with block length, and they
  therefore exhibit an error floor at moderate-to-high signal to noise ratios,
  asymptotically good codes usually converge further away from channel
  capacity. In this paper, we introduce the concept of tuned turbo codes, a family of
  asymptotically good hybrid concatenated code ensembles, where asymptotic minimum distance
  growth rates, convergence thresholds, and code rates can be traded-off using
  two tuning parameters, $\lambda$ and $\mu$.  By decreasing $\lambda$, the
  asymptotic minimum distance growth rate is reduced in exchange for
  improved iterative decoding convergence behavior,
  while increasing $\lambda$ raises the asymptotic minimum distance growth rate at
  the expense of worse convergence behavior, and thus the code
  performance can be tuned to fit the desired application.
  By decreasing $\mu$, a similar tuning behavior can be achieved for
  higher rate code ensembles.
\end{abstract}


%%%%%%%%%%%%%%%%%%%%%%%%%%%%%%%%%%%%%%%%%%%%%%%%%%%%%%%%%%%%%%%%%%%%%%%%%%%%%%%%%%
\section{Introduction} \label{sec:intro}

Turbo codes \cite{BGT93} and multiple parallel concatenated codes (MPCCs) \cite{DP95}
perform very close to the Shannon limit with suboptimum iterative decoding, but the corresponding code
ensembles are asymptotically bad in the sense that their minimum (Hamming)
distance does not grow linearly with block length \cite{KU98}. 
Even the minimum distance of the best code in the ensemble of turbo codes cannot grow more than 
logarithmically with block length \cite{Brei04}. As a
result, their minimum distance may not be sufficient to yield very low
error rates at moderate-to-high signal to noise ratios (SNRs), and
an error floor can occur. 

On the other hand, multiple serially concatenated code (MSCC)
ensembles with three or more component encoders can be
asymptotically good. This has been shown for repeat multiple
accumulate codes in \cite{PfThs03, BMS09, FR09}. There also exist variations 
of standard repeat accumulate codes that are asymptotically good 
\cite{BLS05} but are more complex to encode than classical repeat accumulate codes.

MSCCs in general exhibit good error floor
performance due to their large minimum distance, but they have the
drawback of converging at an SNR further from capacity than parallel
concatenated codes. While the asymptotic distance growth rate of
MSCCs can be made arbitrarily close to the Gilbert-Varshamov Bound
(GVB) by adding more concatenation stages \cite{FR09}, the
iterative decoding convergence behavior of the resulting code
ensembles becomes worse, making codes with more than three
concatenation stages impractical.

An alternative to the above schemes are hybrid concatenated codes (HCCs),
first introduced in \cite{DP97}. They combine the features of
parallel and serially concatenated codes and thus offer more freedom in
code design.
It has been demonstrated in \cite{turbo08} that HCCs can be
designed that perform closer to capacity than MSCCs while still
maintaining a minimum distance that grows linearly with block length.
In particular, small memory-one component encoders are sufficient to yield
asymptotically good code ensembles for such schemes.
The resulting codes provide low complexity encoding and
decoding, and, in many cases, can be decoded using relatively few
iterations.
In \cite{GiAR09}, the analysis of MSCCs and HCCs was extended to the binary erasure channel, 
and stopping set enumerators for the HCCs in \cite{turbo08} were derived.

The HCCs presented in \cite{turbo08} consist of an outer MPCC serially
concatenated with an inner accumulator.
In this paper, we further
elaborate on this code structure and extend the results of \cite{isita08}
to create a family of codes where the asymptotic minimum distance growth rate
and the convergence threshold can be adjusted by varying a
tuning parameter $\lambda$.
In particular, we replace a fraction $1-\lambda$ of the bits at
the output of the inner accumulator with bits taken from the
output of the outer MPCC (see Fig.~\ref{fig:encoder_tuned}).
This leads to a smaller asymptotic distance growth rate for decreasing
$\lambda$ but also to a better iterative decoding threshold.
The resulting code ensembles remain asymptotically good over the range
of all positive values of $\lambda$. We call this family of codes
\textit{tuned turbo codes} (TTCs).
Tuning can also be used to vary the rate of the code. 
To this end, we introduce a second parameter $\mu$, $\lambda \leq \mu \leq 1$,
which denotes the fraction of bits that are kept from the combined output 
of the outer MPCC and the inner accumulator (see Fig.~\ref{fig:encoder_tuned}).
Related code structures have also been investigated in \cite{RGiA11}.

An advantage that TTC ensembles typically have over low-density parity
check (LDPC) code ensembles is that tuning does not change
the encoder structure.
The iterative decoding convergence threshold of an LDPC code ensemble, as well as
their asymptotic minimum distance growth rates, are determined by the
degree distribution of the ensemble. To trade off the iterative
decoding convergence threshold and the asymptotic minimum distance growth
rate, one must vary the degree distribution,
which in general results in a different encoder.

For LDPC codes, quasi-cyclic code constructions are preferred in practice since they 
can be encoded using a low complexity shift register encoder. The quasi-cyclic 
subensembles of LDPC codes, however, are not asymptotically good, since their minimum 
distances are upper bounded by a constant as the circulant sizes increase \cite{SV12}. 

We note that the error floor performance of turbo-like codes with iterative decoding is 
greatly influenced but not solely determined by the minimum Hamming distance of the code, 
the subject of this paper, since pseudo-codewords and trapping sets also play a role.
In Section \ref{sec:encoder}, we present a general encoder structure for TTCs and discuss 
the relevance of the minimum Hamming distance to designing codes with good error floor 
performance. We also introduce four specific types of TTCs that are the focus of our analysis 
throughout the remainder of the paper.
In Section \ref{sec:wefs} we introduce ensemble-average weight 
enumerators for TTCs and their asymptotic expressions.
In Section \ref{sec:Dmin}, the ensemble average weight enumerators are used to bound the 
minimum distance for TTCs, and we present asymptotic minimum distance growth
rates of TTCs for different values of $\lambda$ and $\mu$. Also, a finite length minimum 
distance analysis is performed and the results are shown to be in agreement with the 
asymptotic results.
Section \ref{sec:exit} computes iterative decoding thresholds for 
TTCs using EXIT-charts, and
Section \ref{sec:Tuning} combines the results of the previous two sections
and addresses the tuning behavior of the code constructions. 
Finally, Section \ref{sec:Sim} presents some simulation results, and 
Section \ref{sec:Concl} concludes the paper.

%%%%%%%%%%%%%%%%%%%%%%%%%%%%%%%%%%%%%%%%%%%%%%%%%%%%%%%%%%%%%%%%%%%%%%%%%%%%%%%%%%
\section{Encoder Structure} \label{sec:encoder}

\begin{figure}[t]
    \centering
  \includegraphics[width=1\columnwidth]{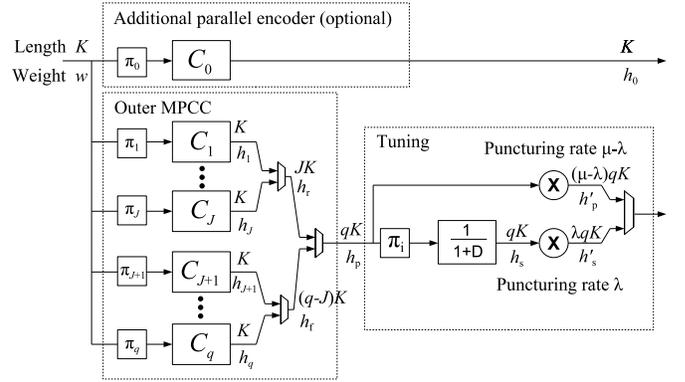}
    \caption{General encoder structure for TTCs with feedforward and recursive convolutional component encoders.}
    \label{fig:encoder_tuned}
\end{figure}

The general structure of the proposed tuned turbo codes is shown in
Fig.~\ref{fig:encoder_tuned}.
They consist of an outer MPCC serially concatenated with an inner rate-1 accumulator 
and optionally an additional rate-1 parallel encoder $\calC_0$.
The outer MPCC consists of a total of $q$ rate-1 component encoders, 
$\calC_1, \calC_2, \ldots, \calC_q$, of which the first $J$ encoders, $3\leq J \leq q$,
are recursive convolutional encoders (RCEs). The remaining $q-J$ encoders in the 
outer MPCC are feedforward convolutional encoders (FFCEs).
(We note that, while in practice it is not necessary 
to precede all component encoders by an interleaver, doing so simplifies the 
analysis and does not change the properties of the code ensemble.)
We denote the combined output weight of the RCEs by $h_{\mathrm{r}}=\sum_{i=1}^J h_i$ 
and the combined output weight of the FFCEs by $h_{\mathrm{f}}=\sum_{i=J+1}^q h_i$, 
where $h_i$ is the output weight of encoder $\calC_i$ in the outer MPCC. The total 
output weight of the outer MPCC $h_{\mathrm{p}}$ is given by 
$h_{\mathrm{p}} = h_{\mathrm{r}} + h_{\mathrm{f}} = \sum_{i=1}^q h_i$.

The output of the outer MPCC enters the serially concatenated inner accumulator, 
whose output weight is denoted by $h_{\mathrm{s}}$. Both the output of the outer 
MPCC and the output of the inner accumulator are punctured, then multiplexed together 
and passed to the channel.
The puncturing rates $(\mu-\lambda)$ and $\lambda$ in Fig.~\ref{fig:encoder_tuned} 
denote the fraction of bits that \emph{survive} after puncturing the outer MPCC and 
the inner accumulator, and $h_p'$ and $h_s'$ are the corresponding weights, respectively.
Finally, $h = h_{\mathrm{s}}' + h_{\mathrm{p}}' + h_0$ represents the total output 
codeword weight.

The parameter $\mu$ is used to control the rate of the TTC ensemble, i.e., considering 
the multiplexed output of the outer MPCC and inner accumulator, a total fraction of 
$\mu$ bits survive puncturing. The rate of the overall ensemble is thus given by
\begin{equation} \label{eq:rate}
    R = \frac{K}{N} = \frac{1}{\mu q + \mathcal{I}_0},
\end{equation}
where $K$ is the input length, $N$ is the total output length,
and $\mathcal{I}_0=1$ if there is an additional parallel encoder and $\mathcal{I}_0=0$ otherwise.
As additional parallel encoders we consider FFCEs or simply a systematic branch.

Tuning the asymptotic minimum distance growth rate and the iterative decoding convergence 
threshold is done by varying the puncturing rate $\lambda$, i.e., changing the fraction 
of bits that come from the output of the inner serially concatenated accumulator. For 
$\lambda=0$, all the bits of the inner accumulator are punctured and the output is the 
(possibly punctured) output of the MPCC. For $\lambda=\mu$ on the other hand, all output 
bits of the tuning section stem from the inner accumulator and none of the bits of the 
outer MPCC survive puncturing.

\begin{figure*}[t]
  \centering
  \includegraphics[width=2\columnwidth]{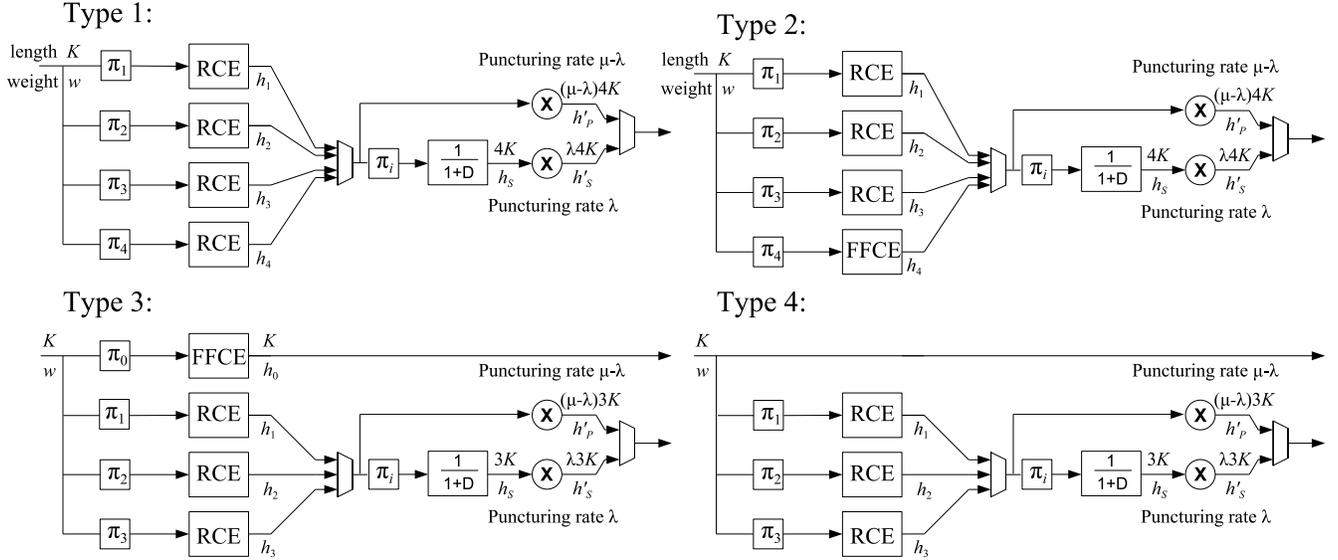}
    \caption{Encoder structure for different TTC types with feedforward and recursive convolutional component encoders (FFCEs and RCEs, respectively).}
    \label{fig:tuned_types}
\end{figure*}

%-----
In the sections of the paper that feature numerical results, 
from Subsection \ref{sub:numDist} onwards,
we consider four different types of TTCs, which are depicted in Fig.~\ref{fig:tuned_types}.
For each type, we consider a version with only 2-state component
encoders and a version with 4-state RCEs in the outer MPCC.
All types are based on the rate $R=1/4$ HCCs introduced in
\cite{turbo08} and, for $\lambda=\mu=1$, are identical to the HCC
in \cite{turbo08}, while for $\lambda=0$, we obtain the
(possibly punctured) outer MPCC plus the optional parallel encoder.

The type 1 and 2 TTC ensembles have a rate $R=1/4$ outer MPCC
with no additional parallel encoder $\calC_0$ (see
Fig.~\ref{fig:tuned_types}).
While the outer MPCC of the type 1 ensemble consists
of four identical rate-1 RCEs, the last encoder $\calC_4$ of the MPCC in the
type 2 ensemble is the 2-state FFCE having generator $[3]_8$ (in octal notation).
The type 3 and 4 TTC ensembles have a rate $R=1/3$ outer MPCC consisting of three 
identical rate-1 RCEs plus an additional parallel encoder $\calC_0$.
In the type 3 ensemble, $\calC_0$ is the $[3]_8$ FFCE,
while in the type 4 ensemble it is simply a systematic branch.
Thus, for $\mu=1$ and $\lambda=0$, when the output of the outer MPCC is not punctured 
and all bits from the inner encoder are punctured, the type 2 and 3 code ensembles are 
identical, while they differ for all other values of $\mu$ and $\lambda$.
The outer MPCC of the type 2 ensemble (with 2-state encoders) was introduced in \cite{MaC01} 
and exhibits excellent iterative decoding behavior due to the presence of the FFCE 
(see \cite{Ching02}).
In all the cases considered in this paper, the 2-state rate-1 RCEs are
accumulators with generator $[1/3]_8$ and the 4-state RCEs are chosen to have the 
generator $[5/7]_8$.

%-----
We decode TTCs iteratively, in a component code oriented fashion, which is a generalization 
of the turbo-decoding principle applied in \cite{BGT93}. Component decoders employ maximum 
a posteriori probability (MAP) decoding strategies and the extrinsic information of one 
component decoder becomes a priori information for the other decoders. In our simulations 
we assume a straightforward iteration schedule, where each component decoder is activated 
once per iteration.

Since we use iterative decoding and not a MAP decoder for the overall code, the performance 
of the decoder in the moderate-to-high SNR region of the additive white Gaussian noise (AWGN) 
channel is greatly influenced but not solely determined by the minimum Hamming distance of 
the code. Pseudo-codewords and trapping sets also play a role in the error floor performance 
of the decoder (see, e.g. \cite{KGKC09, GiAR09}).

TTC ensembles with 2-state component encoders are closely related to LDPC codes and can also 
be decoded using the sum-product algorithm \cite{KFL01}, so it is likely that the pseudo-weight 
properties of TTCs are similar to those of LDPC codes.
In \cite{VK05} it was shown that the minimum AWGN channel pseudo-weight of regular LDPC codes 
grows at best sub-linearly with block length, even though the minimum Hamming distance grows 
linearly with block length. There exist, however, specially constructed code ensembles where 
the minimum binary symmetric channel pseudo-weight can grow linearly with block length \cite{FMSSW07}. 

Since the minimum Hamming distance is an upper bound on the minimum pseudo-weight, we expect 
that designing TTC ensembles whose minimum distance grows linearly with block length will 
lead to code ensembles that also possess good pseudo-weight properties. This expectation is 
supported by the finite length (Hamming) distance analysis in Section~\ref{sec:Dmin} and the 
simulation results in Section~\ref{sec:Sim}, both of which show that code ensembles with large 
minimum Hamming distance exhibit low error floors.

%%%%%%%%%%%%%%%%%%%%%%%%%%%%%%%%%%%%%%%%%%%%%%%%%%%%%%%%%%%%%%%%%%%%%%%%%%%%%%%%%%
\section{Preliminaries} \label{sec:wefs}

\subsection{Weight Enumerators}

The weight spectrum of an $(N,K)$ linear encoder $\calC(N)$
is described by its weight enumerator (WE) $A^{\calC(N)}_{h}$,
which specifies the number of codewords with output weight $h$.
Likewise, let $A^{\calC(N)}_{w,h}$ denote the input-output weight enumerator (IOWE),
which specifies the number of codewords with input weight $w$ and
output weight $h$.
To investigate the distance properties of tuned turbo code ensembles,
we consider the ensemble average of the above quantities.
For an encoder ensemble $\mathbf{C}(N)$ of length $N$, we write the average IOWE as
\begin{equation}
    \bar{A}^{\mathbf{C}(N)}_{w,h} = \frac{1}{|\mathbf{C}(N)|}
    \sum_{\calC(N) \in \mathbf{C}(N)} A^{\calC(N)}_{w,h},
\end{equation}
where $|\mathbf{C}(N)|$ denotes the size of $\mathbf{C}(N)$.
When the members of $\mathbf{C}(N)$ are equally likely, we obtain the average WE as
\begin{equation} \label{eq:we}
    \bar{A}^{\mathbf{C}(N)}_{h} = \sum_{w=1}^{K} \bar{A}^{\mathbf{C}(N)}_{w,h}.
\end{equation}
The average WE represents the expected number of codewords of weight $h$ if a code is randomly
chosen from the ensemble $\mathbf{C}(N)$.
In the rest of the paper, whenever the context is clear, we will omit the parameter $N$.

To obtain the average WE $\bar{A}^{\mathbf{C}_{\mathrm{TTC}}(N)}_{h}$ of 
TTC ensembles, we use the uniform
interleaver analysis introduced in \cite{BDMP98a}. The uniform interleaver is a
probabilistic device that maps an input block of weight $w$ and length $K_\calC$
into all its possible $\binom{K_\calC}{w}$ permutations with equal probability,
thus decoupling the component encoders in a concatenated code and creating a code
ensemble with equally likely members.
An $(N_\calC,K_\calC)$ component encoder $\calC(N_\calC)$, preceded by a uniform 
interleaver, results in the input-output weight distribution (IOWD)
\begin{equation} \label{eq:iowd}
    \mathbb{P}^{\calC(N_\calC)}_{w,h} = 
    \frac{A^{\calC(N_\calC)}_{w,h}}{\binom{K_\calC}{w}},
\end{equation}
where $\mathbb{P}^{\calC(N_\calC)}_{w,h}$ is the probability that encoder $\calC(N_\calC)$
transforms an input of weight $w$ into an output of weight $h$.
For $(N_\calC,N_\calC)$ 2-state component encoders, the IOWE can be given in closed
form as \cite{DJM98}
\begin{equation}\label{eq:WEacc}
        {A}_{w,h}^{\mathrm{Acc}(N_\calC)} 
        = {A}^{\mathrm{FF}(N_\calC)}_{h,w}
        = { \binom{N_\calC-h}{\left\lfloor w/2 \right\rfloor}
        \binom{h-1}{\left\lceil w/2\right\rceil-1}},
\end{equation}
where $w/2 \leq h \leq N_\calC-w/2$, ``Acc'' represents the accumulator, and ``FF'' 
represents the 2-state FFCE with generator $[3]_8$.

In the case of TTCs, component codes may be punctured.
The IOWE of punctured accumulators was analyzed in \cite{ADY07}
by considering the serial concatenation of an accumulator and a
single-parity-check. Using this approach, only regular puncturing patterns and
puncturing rates of $\lambda = 1/i$ with $i \in \mathbb{N}$ can be realized.
To be able to vary $\lambda$ continuously, we therefore consider
random puncturing, and the code ensembles we analyze
are formed over all interleaver realizations, as well as over all
possible puncturing patterns.
Using random puncturing, the probability that a codeword of length $N$
and weight $h$ before puncturing is punctured to a codeword of
length $N'=\lambda N$ and weight $h'$ is given by the hypergeometric distribution
\begin{equation}
    \mathbb{P}^{\mathrm{P}(N')}_{h,h',\lambda} =
    \frac{\binom{N'}{h'} \binom{N-N'}{h-h'}}{\binom{N}{h}},
\label{eq:WE_punct}
\end{equation}
where \eqref{eq:WE_punct} represents the IOWD of the random puncturing operation 
and we require $h' \leq N'$, $h-h' \leq N-N'$, and $h' \leq h$.
Throughout the paper we define the binomial coefficient $\binom{n}{k}$
to be zero if $n<k$.

The average component input-output weight enumerator (CIOWE) of an $(N,K)$ TTC, 
$\bar{A}_{w,h_0,h_1,\ldots,h_q,h_{\mathrm{s}},h_{\mathrm{p}}',h_{\mathrm{s}}'}^{\mathbf{C}_{\mathrm{TTC}}(N)}$, 
is the average number of codewords with fixed input and output weights 
$w,h_0,h_1,\ldots,h_q,h_{\mathrm{s}},h_{\mathrm{p}}',$ and $h_{\mathrm{s}}'$ 
of each component encoder in the TTC.
The CIOWE is simply the product of 
the IOWDs of the components times the number of permutations of the input sequence, i.e.,
\begin{equation} \label{eq:cwef_tt}
\begin{aligned}
&\bar{A}_{w,h_0,h_1,\ldots,h_q,h_{\mathrm{s}},h_{\mathrm{p}}',h_{\mathrm{s}}'}^{\mathbf{C}_{\mathrm{TTC}}(N)} =\\ 
                &\binom{K}{w}  \left(
                \prod_{i=0}^{q} \mathbb{P}_{w,h_i}^{\calC_i(K)} \right)
                \cdot \mathbb{P}_{h_{\mathrm{p}},h_{\mathrm{s}}}^{\mathrm{Acc}(qK)}
                \cdot \mathbb{P}_{h_{\mathrm{p}},h_{\mathrm{p}}',\mu-\lambda}^{\mathrm{P}(qK(\mu-\lambda))}
                \cdot \mathbb{P}_{h_{\mathrm{s}},h_{\mathrm{s}}',\lambda}^{\mathrm{P}(qK\lambda)},
\end{aligned}
\end{equation}
where we denote the total output weight of the outer MPCC as $h_{\mathrm{p}}= \sum_{i=1}^{q}h_i$.
If there is no additional parallel encoder, we define
$\mathbb{P}^{\calC_0(K)}_{w,h_0}$ to be one for $h_0=0$ and zero otherwise.

The ensemble average IOWE of a TTC, $\bar{A}_{w,h}^{\mathbf{C}_{\mathrm{TTC}}(N)}$,
is then the summation over all CIOWEs such that the codeword has weight $h$.
To include the total output weight $h$ in the CIOWE, we represent
the punctured weight of the inner serial accumulator as $h_{\mathrm{s}}' = h-h_{\mathrm{p}}'-h_0$,
thus obtaining the IOWE
\begin{equation} \label{eq:wef_tt}
\begin{aligned}
 &\bar{A}_{w,h}^{\mathbf{C}_{\mathrm{TTC}}(N)} = \\
      &\sum_{h_0=\mathcal{I}_0}^{K} 
      \sum_{h_1=1}^{K} \cdots
   		\sum_{h_q=1}^{K} \sum_{h_{\mathrm{s}}=1}^{qK}
   		\sum_{h_{\mathrm{p}}'=0}^{h}
	\bar{A}_{w,h_0,h_1,\ldots,h_q,h_{\mathrm{s}},h_{\mathrm{p}}',h}^{\mathbf{C}_{\mathrm{TTC}}(N)}.
\end{aligned}
\end{equation}
Note that, with random puncturing, it is possible that all the weight is punctured
and therefore the enumeration of punctured weights starts at zero.

%%%%%%%%%%%%%%%%%%%%%%%%%%%%%%%%%%%%%%%%%%%%%%%%%%%%%%%%%%%%%%%%%%%%%%%%%%%%%%
\subsection{The Spectral Shape} \label{sec:SpShape}

To investigate the asymptotic minimum distance properties of tuned
turbo codes as the block length $N$ tends to infinity, we will make use 
of the asymptotic spectral shape function originally introduced by 
Gallager \cite{Gal63},
\begin{equation}\label{eq:specshape}
        r(\rho) = \limsup_{N\rightarrow\infty} \frac{\ln \bar{A}^{\mathbf{C}(N)}
        _{\left\lfloor \rho N \right\rfloor}}{N},
\end{equation}
where $\rho= \frac{h}{N}$ is the normalized codeword weight.
The spectral shape is the exponential part of the average WE normalized by the block length $N$. 
When $r(\rho)<0$, the average number of codewords with normalized weight $\rho$ goes exponentially 
to zero as $N$ tends to infinity. When $r(\rho)>0$, on the other hand, the average number of 
codewords with normalized weight $\rho$ grows exponentially in $N$. When $r(\rho)=0$, the average 
number of codewords with normalized weight $\rho$ does not exhibit exponential growth --- it might 
increase or decrease polynomially, for example.

Similarly, we define the asymptotic IOWD of an $(N_\calC,K_\calC)$ 
component code $\calC(N_\calC)$ as
\begin{equation} \label{eq:def_f}
    f^{\calC}_{\alpha, \beta} = \lim_{N_\calC \to \infty} \frac{\ln
        \mathbb{P}_{\left\lfloor \alpha K_\calC \right\rfloor,
        \left\lfloor \beta N_\calC \right\rfloor}^{\calC(N_\calC)}}{N_\calC},
\end{equation}
where $\alpha$ and $\beta$ are the normalized input and output weight
w.r.t. the input block length $K_\calC$ and the output block length $N_\calC$,
respectively, of code $\calC(N_\calC)$.
Stirling's approximation can be used to bound the binomial coefficients as
\begin{equation} \label{eq:stirling}
    \frac{e^{n\H{k/n}}}{n+1} \leq \binom{n}{k} \leq e^{n\H{k/n}},
\end{equation}
where $\H{x}=-x\ln x - (1-x)\ln(1-x)$ denotes the binary entropy
function using the natural logarithm.

Using \eqref{eq:iowd}, \eqref{eq:WEacc}, \eqref{eq:def_f}, and \eqref{eq:stirling}
the asymptotic IOWD of the accumulator is given by
\begin{equation}
\label{eq:AccAIOWD}
        f^{\mathrm{Acc}}_{\alpha, \beta} = (1-\beta) \H{\frac{\alpha}{2(1-\beta)}}
                        + \beta \H{\frac{\alpha}{2 \beta}} - \H{\alpha},
\end{equation}
where $\alpha = w/N_\calC$ and $\beta = h/N_\calC$ (see also \cite{PfThs03}). In the 
same way, the asymptotic IOWD of the 2-state FFCE is given by
\begin{equation}
\label{eq:1pDAIOWD}
        f^{\mathrm{FF}}_{\alpha, \beta} = (1-\alpha) \H{\frac{\beta}{2(1-\alpha)}}
                        + \alpha \H{\frac{\beta}{2 \alpha}} - \H{\alpha}.
\end{equation}
Similarly, by using \eqref{eq:WE_punct}, \eqref{eq:def_f}, and \eqref{eq:stirling},
the asymptotic IOWD of the random puncturing operation is given by
\begin{equation}
\label{eq:PunctAIOWD}
        f^{\mathrm{P}}_{\beta, \beta', \lambda} =
            \H{\beta'} + \frac{1-\lambda}{\lambda} 
            		\H{\frac{\beta-\lambda \beta'}{1-\lambda}}
             - \frac{1}{\lambda}\H{\beta},
\end{equation}
where $\beta = h/N_\calC$ and $\beta' = h'/(\lambda N_\calC)$.

We now define the asymptotic CIOWE of an $(N,K)$ TTC as
\begin{equation} \label{eq:ACIOWE}
    \mathcal{F}_{\alpha,\rho_0,\rho_1,\ldots,\rho_q,\rho_{\mathrm{s}}, \rho_{\mathrm{p}}',\rho}
    				^{\mathbf{C}_{\mathrm{TTC}}} = \lim_{N\rightarrow\infty}
        \frac{ \ln \bar{A}_{w,h_0,h_1,\ldots,h_q,h_{\mathrm{s}},h_{\mathrm{p}}',h}^{\mathbf{C}_{\mathrm{TTC}}(N)} } {N},
\end{equation}
where $\alpha = w/K$, $\rho_i = h_i/K$, $i=0,1,\ldots,q$, $\rho_{\mathrm{s}} = h_{\mathrm{s}} / (qK)$,
$\rho_{\mathrm{p}}' = h_{\mathrm{p}}'/(q(\mu-\lambda)K)$, and $\rho = h/(q\mu + \mathcal{I}_0)K = h/N$.
Using \eqref{eq:ACIOWE}, we rewrite the asymptotic spectral shape as the 
optimization problem
\begin{equation}\label{eq:specshape2}
        r(\rho) = \sup_{\alpha,\rho_0,\ldots,\rho_{\mathrm{p}}'}
            \mathcal{F}_{\alpha,\rho_0,\rho_1,\ldots,\rho_q,\rho_{\mathrm{s}},\rho_{\mathrm{p}}',\rho}^{\mathbf{C}_{\mathrm{TTC}}}.
\end{equation}

We obtain the asymptotic CIOWE of a TTC by inserting its CIOWE
\eqref{eq:cwef_tt} into \eqref{eq:ACIOWE}.
The logarithm transforms the product of IOWDs in \eqref{eq:cwef_tt} into a sum, and in the limit
as the block lengths of each component code $N_{\calC_i}$ tend to infinity,
the asymptotic CIOWE of a TTC can be written
in terms of the asymptotic IOWDs of its component codes, weighted
by their respective block lengths divided by $N$, as
\begin{equation} \label{eq:ACIOWE2}
\begin{aligned}
    & \mathcal{F}_{\alpha,\rho_0,\rho_1,\ldots,\rho_q,\rho_{\mathrm{s}},\rho_{\mathrm{p}}',\rho}
    				^{\mathbf{C}_{\mathrm{TTC}}} = \\
    	&	qR \left(	\frac{1}{q}\H{\alpha} + \frac{1}{q} \sum_{i=0}^q f^{\calC_i}_{\alpha,\rho_i} +
                     	f^{\mathrm{Acc}}_{\rho_{\mathrm{p}}, \rho_{\mathrm{s}}} + \right.\\
            & \left. (\mu-\lambda) f^{\mathrm{P}}_{\rho_{\mathrm{p}}, {\rho'}_{\mathrm{p}}, (\mu-\lambda)} +
                     \lambda f^{\mathrm{P}}_{ \rho_{\mathrm{s}}, \rho_{\mathrm{s}}', \lambda} \right),		
\end{aligned}
\end{equation}
where $\rho_{\mathrm{p}}=h_{\mathrm{p}}/qK$ and
$R=K/N$ is the rate of the TTC given by \eqref{eq:rate}.
To include the normalized total output weight $\rho=h/N$ in the asymptotic CIOWE, 
we represent the normalized punctured weight of the inner serial accumulator
$\rho_{\mathrm{s}}'$ as
\begin{equation}
\begin{aligned}
    \rho_{\mathrm{s}}' = & \, \frac{h_{\mathrm{s}}'}{q\lambda K} = 
    	\frac{h - h_{\mathrm{p}}' - h_0}{q\lambda K} \\
    	= & \frac{\rho/R-q(\mu-\lambda){\rho_{\mathrm{p}}}'- \rho_0}{q\lambda}.
\end{aligned}
\end{equation}

If there is some $\hat{\rho} > 0$ such that $r(\rho) < 0$ for all $0 < \rho < \hat{\rho}$, 
we would immediately have that $\hat{\rho}$ is the asymptotic growth rate of the minimum 
distance of the ensemble.
However, this is not the case for TTCs.
%-------
\begin{prop} \label{prop:zero}
	For $0 \leq \rho < Rq\lambda$, the spectral shape of a TTC cannot be negative but is 
	lower bounded by $r(\rho)=0$.
\end{prop}
\begin{IEEEproof}
	The proposition is trivially proved by setting $\rho_{\mathrm{p}} = 0$ (which implies 
	$\alpha=0$, $\rho_0=0$, $\rho_i=0$ for $i=1,\ldots,q$, and $\rho_{\mathrm{p}}'=0$) and	
	$\rho_{\mathrm{s}} = \rho_{\mathrm{s}}' = \rho /({Rq\lambda})$. Setting $\rho_{\mathrm{p}} = 0$
	results in the asymptotic IOWDs of the 2-state component encoders \eqref{eq:AccAIOWD},
	 \eqref{eq:1pDAIOWD} in the CIOWE of a TTC to be zero and setting
	 $\rho_{\mathrm{s}} = \rho_{\mathrm{s}}'$ results in the asymptotic IOWD of the 
	 random puncturing operation \eqref{eq:PunctAIOWD} to be zero, resulting in
	\begin{equation*}
	\mathcal{F}_{0,0,0,\ldots,0,(\rho/({Rq\lambda})),
	0,\rho}^{\mathbf{C}_{\mathrm{TTC}}}=0.
	\end{equation*}
\end{IEEEproof}
%-------
Thus, it cannot be directly concluded that the resulting ensembles are asymptotically good, 
but we will show in the next section that the 2-state ensembles are indeed asymptotically good
and we conjecture the same for the 4-state ensembles.

%%%%%%%%%%%%%%%%%%%%%%%%%%%%%%%%%%%%%%%%%%%%%%%%%%%%%%%%%%%%%%%%%%%%%%%%%%%
\section{Minimum Distance Analysis}
\label{sec:Dmin}

In this section we make use of the expressions from the previous section
to perform both an asymptotic and a finite length minimum distance analysis
of tuned turbo codes with 2-state component encoders.

%-----------------------------------------------------------------
\subsection{Asymptotic Analysis} \label{sub:AsDist}

For a TTC with $q$ encoders and $J$ accumulators in the outer MPCC
(see Fig.~\ref{fig:encoder_tuned}),
the probability that a randomly chosen code from the ensemble
has minimum distance $d_\mathrm{min} < d$ is upper bounded by
\begin{align} 
       & \P{d_\mathrm{min} <  d} \leq  (\bar{A}_{0}^{\mathbf{C}_{\mathrm{TTC}}(N)}-1)+\sum_{h=1}^{d-1}\bar{A}_{h}^{\mathbf{C}_{\mathrm{TTC}}(N)}
                \label{eq:dminBound} \\
            & = \sum_{w=1}^{K} \sum_{h_0=\calI_0}^{K} 
             			\sum_{h_1=1}^{K} \cdots
             		  \sum_{h_q=1}^{K} \sum_{h_{\mathrm{s}}=1}^{Kq}
             		  \sum_{h_{\mathrm{p}}'=0}^{\lfloor Kq(\mu-\lambda) \rfloor} \sum_{h=0}^{d-1}
                    \bar{A}_{w,h_0,h_1,\ldots,h_q,h_{\mathrm{s}},h_{\mathrm{p}}', h}^{\mathbf{C}_{\mathrm{TTC}}(N)}.
                    \label{eq:dminBound2}
\end{align}
Note that, while the average number of all-zero codewords $\bar{A}_0^{\mathbf{C}(N)}$ 
equals $1$ for unpunctured linear codes, with punctured codes there is a possibility 
that all the weight is removed by the puncturing operation. We take the probability of 
this event into account with the term $(\bar{A}_{0}^{\mathbf{C}_{\mathrm{TTC}}(N)}-1)$ 
in \eqref{eq:dminBound} and with the summation over $h$ starting from zero in \eqref{eq:dminBound2}, 
while the summation over $w$ starts at $w=1$.

We define the value $\hat{\rho}$ as follows.
\begin{defn} \label{def:rhohat}
	Let $0 \leq \hat{\rho} < Rq\lambda/2$ be such that, for all $0 \leq \rho < \hat{\rho}$, 
	the unique supremum of the asymptotic CIOWE of TTCs given by \eqref{eq:ACIOWE2} is achieved 
	for $\rho_{\mathrm{p}} = 0$ and $\rho_{\mathrm{s}} = \rho_{\mathrm{s}}' = \rho /({Rq\lambda})$.
\end{defn}
%---------------------------------------------
% Floating equation (24)
\begin{figure*}[!b]
\vspace{4pt}
%separator
\hrulefill
\normalsize
\newcounter {MYtempeqncount}
\setcounter{MYtempeqncount}{\value{equation}}
\setcounter{equation}{23}
\begin{equation} \label{eq:a2bound2}
\begin{aligned}
  A_2 \leq & (NRq+1)^{q+5} \exp \left\{ 
          		\max_{\stackrel{h_{\mathrm{p}} > h_{\mathrm{p}}^*, h < \left\lceil N(\hat{\rho}-\epsilon) \right\rceil}
          		{w,h_0,h_{\mathrm{p}}',h_{\mathrm{s}}}}
       				 N \cdot \mathcal{F}^{\calC_{\mathrm{TTC}}}_{\frac{w}{NR},\, \frac{h_0}{NR},\, 
       				 			\frac{h_1}{NRq},\, \ldots,\, \frac{h_q}{NRq},\, \frac{h_{\mathrm{s}}}{NRq},\, 
       				 			\frac{h_{\mathrm{p}}'}{NRq(\mu-\lambda)},\, \frac{h}{N} } 
       				 			+ (2q+8)\ln(NRq+1) \right\} \\
       = & \exp \left\{ 
          		\max_{\stackrel{h_{\mathrm{p}} > h_{\mathrm{p}}^*, h < \left\lceil N(\hat{\rho}-\epsilon) \right\rceil}
          		{w,h_0,h_{\mathrm{p}}',h_{\mathrm{s}}}}
       				 N \cdot \mathcal{F}^{\calC_{\mathrm{TTC}}}_{\frac{w}{NR},\, \frac{h_0}{NR},\, 
       				 			\frac{h_1}{NRq},\, \ldots,\, \frac{h_q}{NRq}
       				 			\frac{h_{\mathrm{s}}}{NRq},\, \frac{h_{\mathrm{p}}'}{NRq(\mu-\lambda)},\,
                    \frac{h}{N} } + (3q+13)\ln (NRq+1) \right\}
\end{aligned}
\end{equation}
\setcounter{equation}{\value{MYtempeqncount}}
\end{figure*}
%------------------------------------------------

Following the procedure established in \cite{JM02} and \cite{FR09}, 
we can split \eqref{eq:dminBound2} into two parts,
$A_1$ and $A_2$, depending on the output weight of the outer MPCC 
$h_{\mathrm{p}}= \sum_{i=1}^q h_i$.
For any positive integer $h_{\mathrm{p}}^*$, $q \leq h_{\mathrm{p}}^* \leq Kq$, we can write
\begin{equation}
\begin{aligned}
	& \P{d_\mathrm{min} < d} = \\
	& \underbrace{ \P{d_\mathrm{min} < d \cap h_{\mathrm{p}} \leq h_{\mathrm{p}}^*} }_{A_1} + 
			\underbrace{ \P{d_\mathrm{min} < d \cap h_{\mathrm{p}} > h_{\mathrm{p}}^*} }_{A_2}.
\end{aligned}
\end{equation}
We now proceed to show that, with appropriately chosen values of $h_{\mathrm{p}}^*$ and $d$,
$A_1 \rightarrow 0$ and $A_2 \rightarrow 0$ as $N \rightarrow \infty$ for all 
$d<\left\lceil N(\hat{\rho}-\epsilon) \right\rceil$, where $\epsilon>0$ is an arbitrarily 
small constant, which implies that the ensemble is asymptotically good
with asymptotic minimum distance growth rate $\hat{\rho}$.

\begin{lemma} \label{lem:a1}
	As $N \rightarrow \infty$, for all $J>2$ and $h_{\mathrm{p}}^* \leq N^{\frac{J-2}{J}-\epsilon}$, 
	we have
	\begin{equation}
		A_1 = \P{d_\mathrm{min} < \left\lceil N(\hat{\rho}-\epsilon) \right\rceil \cap h_{\mathrm{p}} \leq h_{\mathrm{p}}^*} \rightarrow 0
	\end{equation}
	for arbitrarily small values of $\epsilon>0$.
\end{lemma}
\begin{IEEEproof}
	Using the simple upper bound
	\begin{equation*}
	\begin{aligned}
	A_1 = & \P{d_\mathrm{min} < \left\lceil N(\hat{\rho}-\epsilon) \right\rceil \cap h_{\mathrm{p}} \leq h_{\mathrm{p}}^*}\\
	      & \leq \P{h_{\mathrm{p}} \leq h_{\mathrm{p}}^*},
	\end{aligned}
	\end{equation*}
	the problem is reduced to finding the asymptotic minimum distance of an MPCC 
	with $J$ parallel concatenated RCEs, which was lower bounded in \cite{KU98} and \cite{JM02} as
	\begin{equation*} \label{eq:dminMPCC}
    \mathbb{P}\left(h_{\mathrm{p}} \leq N^{\frac{J-2}{J}-\epsilon} \right) \leq C_1 \cdot N^{-\epsilon/2}
	\end{equation*}
	for arbitrarily small values of $\epsilon>0$, some positive constant $C_1$,
	and $N$ sufficiently large.
\end{IEEEproof}

Now, considering $A_2$, we upper bound the $q+5$ sums
in \eqref{eq:dminBound2} by their maximum element times $NRq+1$, 
which is an upper bound on the number of terms in each sum, and we obtain
\begin{equation} \label{eq:a2bound}
\begin{aligned}
		A_2 = & \P{d_\mathrm{min}< \left\lceil N(\hat{\rho}-\epsilon) \right\rceil \cap h_{\mathrm{p}}>h_{\mathrm{p}}^*} \\
     \leq & (NRq+1)^{q+5} \max_{\substack{
     		\sum_{i=1}^q h_i = h_{\mathrm{p}} > h_{\mathrm{p}}^* \\
     		h < \left\lceil N(\hat{\rho}-\epsilon) \right\rceil \\
     		w,h_0,h_{\mathrm{p}}',h_{\mathrm{s}}} }
       		\bar{A}_{w,h_0, h_1, \ldots, h_q, h_{\mathrm{s}},h_{\mathrm{p}}', h}^{\mathbf{C}_{\mathrm{TTC}}(N)}.
\end{aligned}
\end{equation}

Using Stirling's approximation \eqref{eq:stirling}, we can upper bound 
each of the $q+4$ IOWDS in the CIOWE of \eqref{eq:cwef_tt} as
\begin{equation*}
	\mathbb{P}_{w_i,h_i}^{\calC_i(N_{\calC_i})} \leq \exp \left\{ 
				N_{\calC_i} f_{\alpha_i,\beta_i}^{\calC_i} + 2\ln(N_{\calC_i}+1)  \right\}.
\end{equation*}
Then using the notation of the asymptotic CIOWE \eqref{eq:ACIOWE2} and
upper bounding $N_{\calC_i}+1$ by $NRq+1$, we obtain \eqref{eq:a2bound2}.
%---------------------------------------------
% Original place of floating equation (24)
\setcounter{equation}{24}
%---------------------------------------------

Thus, to bound $A_2$ it is necessary to examine the asymptotic CIOWE and the asymptotic 
spectral shape \eqref{eq:specshape2} of TTCs.
We now show that if there exists a $\hat{\rho}>0$ as defined in Definition~\ref{def:rhohat}, 
we have $A_2 \rightarrow 0$ as $N \rightarrow \infty$.
To this end we make use of the log-concavity of the IOWDs of the component encoders.
\begin{prop} \label{prop:concave}
It holds that:
	\begin{enumerate}
		\item For a fixed input weight $w$, the IOWD $\mathbb{P}_{w,h}^{\mathrm{Acc}(N_\calC)}$ 
    of the accumulator forms a logarithmically concave sequence in the output 
    weight $h$ and its maximum occurs at $h=N_\calC/2$.
    
    \item For a fixed input weight $w$, the IOWD $\mathbb{P}_{w,h}^{\mathrm{FF}}{(N_\calC)}$ 
    of the 2-state FFCE forms a logarithmically concave sequence in the output 
    weight $h$ and its maximum occurs at $h=2w( 1-w/{N_\calC})$.
    
    \item For a fixed input weight $h$, the IOWD $\mathbb{P}_{h,h',\lambda}^{\mathrm{P}(N')}$
    of the random puncturing operation forms a strictly logarithmically 
    concave sequence in the output weight $h'$ and its maximum occurs at $h'=\lambda h$.
	\end{enumerate}
\end{prop}
The proofs of these statements can be found in Appendix~\ref{app:concave}.
From Proposition~\ref{prop:concave} it follows that, for a fixed input weight $w$ 
and a fixed total output weight of the RCEs $h_{\mathrm{r}}=\sum_{i=1}^J h_i$  
(see Fig.~\ref{fig:encoder_tuned}), the CIOWE of TTCs is maximized when the 
RCEs in the outer MPCC contribute equally to $h_{\mathrm{r}}$, i.e., when 
$h_i=h_{\mathrm{r}}/J$, or $\rho_{\mathrm{i}} = 
 h_{\mathrm{r}}/JK = \rho_{\mathrm{r}}$, $i \in \{1,\ldots,J\}$.
Equivalently, the CIOWE of TTCs is maximized when the FFCEs in the outer MPCC
contribute equally to $h_{\mathrm{f}}$, i.e., when 
$h_i=h_{\mathrm{f}}/(q-J)$, or $\rho_i = 
 h_{\mathrm{f}}/(q-J)K = \rho_{\mathrm{f}}$, $i \in \{J+1,\ldots,q\}$.
Thus we can substitute $\rho_{\mathrm{s}}$ and $\rho_{\mathrm{f}}$ for the $\rho_i$ 
in the asymptotic CIOWE and the number of variables in the maximization problem in 
\eqref{eq:a2bound2} is reduced.
The normalized output weight of the outer MPCC is then given by 
$\rho_{\mathrm{p}} = (J\rho_{\mathrm{s}}/q + (q-J)\rho_{\mathrm{f}}/q)$.

%--------------
\begin{lemma} \label{lem:a2}
	If there exists a $\hat{\rho}>0$ then for $J>2$ RCEs in the outer MPCC and 
	\begin{equation*}
		\lim_{n\rightarrow \infty} \frac{\ln (NRq+1)}{h^*_{\mathrm{p}}}=0
	\end{equation*}
	we have
	\begin{equation*}
		A_2 = \P{d_\mathrm{min}< \left\lceil N(\hat{\rho}-\epsilon) \right\rceil \cap h_{\mathrm{p}}>h_{\mathrm{p}}^*} \rightarrow 0
	\end{equation*}
	as $N \rightarrow \infty$, where $\epsilon>0$ is an arbitrarily small constant.
\end{lemma}
%-------------
\begin{IEEEproof}
	We investigate the asymptotic CIOWE \eqref{eq:ACIOWE2} in the region $0 \leq \rho < \hat{\rho}$
	by splitting it into two parts, $RqF_1$ and $RqF_2$,
and write
\begin{equation} \label{eq:F1F2}
	N \cdot \mathcal{F}^{\calC_{\mathrm{TTC}}}_{\frac{w}{NR},\, \frac{h_0}{NR},\, 
       				 			\frac{h_{\mathrm{s}}}{NRJ},\, \frac{h_{\mathrm{f}}}{NR(q-J)},\, \frac{h_{\mathrm{r}}}{NRq},\, 
       				 			\frac{h_{\mathrm{p}}'}{NRq(\mu-\lambda)},\, \frac{h}{N} } = NRq (F_1 + F_2)
\end{equation}
with
\begin{equation}
\begin{aligned}
		F_1 = & \frac{1}{q}\H{ \frac{w}{NR} } 
         		+ \frac{\calI_0}{q} f^{\calC_0}_{\frac{w}{NR}, \, \frac{h_0}{NR} } 
            + \frac{J}{q} f^{\mathrm{Acc}}_{\frac{w}{NR},\, \frac{h_{\mathrm{r}}}{NRJ} } \\
           & + \frac{q-J}{q} f^{\mathrm{FF}}_{\frac{w}{NR},\, \frac{h_{\mathrm{f}}}{NR(q-J)} }
            + f^{\mathrm{Acc}}_{\frac{h_{\mathrm{p}}}{NRq},\, \frac{h_{\mathrm{s}}}{NRq}} \label{eq:F1}
\end{aligned}
\end{equation}
\begin{equation}
    F_2 = (\mu-\lambda) f^{\mathrm{P}}_{\frac{h_{\mathrm{p}}}{NRq} ,\, 
        						\frac{h_{\mathrm{p}}'}{NRq(\mu-\lambda)},\, \mu-\lambda }
            + \lambda f^{\mathrm{P}}_{ \frac{h_{\mathrm{s}}}{NRq},\, \frac{h-h_{\mathrm{p}}'-h_0}{NR\lambda q},\,
             \lambda}.
             \label{eq:F2}
\end{equation}
The term $F_1$ includes the asymptotic IOWDs of the encoders, whereas $F_2$ 
includes the asymptotic IOWDs of the random puncturing operation.
From Definition~\ref{def:rhohat} we have that, for $0 \leq \rho < \hat{\rho}$, 
the spectral shape has its supremum at $r(\rho)=0$, which is achieved for 
$\rho_p=0$ and $\rho_s=\rho_s'=\rho/(Rq\lambda)$.

First we note that $F_2 \leq 0$, with $F_2 = 0$ for $\rho_s=\rho_s'$ and $F_2<0$ 
otherwise (see the proof of part 3 of Proposition~\ref{prop:concave}), so we can 
simply upper bound the term $NRqF_2$ by zero.

Next we note that  $F_1$ tends to zero for $\rho_{\mathrm{p}} \rightarrow 0$ 
(see the proof of Proposition~\ref{prop:zero}), and hence to upper bound $NRqF_1$ 
we consider all possible $h_{\mathrm{p}}>h_{\mathrm{p}}^*$ such that $\rho_{\mathrm{p}}$ 
tends asymptotically to zero, i.e.,
\begin{equation*}
    \lim_{N \rightarrow \infty} \frac {h_{\mathrm{p}}}{N} = 0.
\end{equation*}
Then we have
\begin{equation} \label{eq:derivative}
    \lim_{\substack{N \rightarrow \infty \\h_{\mathrm{p}}/N \rightarrow 0}} NRq\,F_1 = 
    \lim_{\substack{N \rightarrow \infty \\h_{\mathrm{p}}/N \rightarrow 0}} 
    		h_{\mathrm{p}}\cdot\frac{F_1}{\rho_{\mathrm{p}}} = 
    h_{\mathrm{p}} \cdot \frac{d}{d \rho_{\mathrm{p}}} \left. F_1 \right|_{\rho_{\mathrm{p}}=0},
\end{equation}
where both $F_1$ and $\rho_{\mathrm{p}}=h_{\mathrm{p}}/(NRq)$ tend asymptotically to zero 
as $N \rightarrow \infty$ and the fraction $F_1/\rho_{\mathrm{p}}$ is the difference 
quotient of the point $F_1|_{\rho_{\mathrm{p}}=0}$, which as $N \rightarrow \infty$ yields 
the total derivative of $F_1$ with respect to $\rho_{\mathrm{p}}$ evaluated at $\rho_{\mathrm{p}}=0$.

We show in Appendix~\ref{app:derivative} that, if there exists a $\hat{\rho}>0$, then,
for $J>2$ accumulators in the outer MPCC, the total derivative of $F_1$ with respect 
to $\rho_{\mathrm{p}}$ evaluated at $\rho_{\mathrm{p}}=0$ is bounded by
\begin{equation} \label{eq:der1}
     \frac{d}{d \rho_{\mathrm{p}}} \left. F_1 \right|_{\rho_{\mathrm{p}}=0} \leq -C_2,
\end{equation}
for some positive constant $C_2$ and for $\rho_{\mathrm{s}}<1/2$.

So, for $\rho_{\mathrm{p}}=0$, from \eqref{eq:a2bound2}, 
\eqref{eq:F1F2}, \eqref{eq:F1}, \eqref{eq:F2}, \eqref{eq:derivative}, 
and \eqref{eq:der1} we can write
\begin{equation} \label{eq:A2sum}
    A_2 \leq \exp\left\{ -C_2\cdot h_{\mathrm{p}} + (3q+13)\ln (NRq+1) \right\},
\end{equation}
and $\lim_{N \rightarrow \infty} A_2=0$ for all $h_{\mathrm{p}}^*$ that satisfy
$\lim_{N \rightarrow \infty} \ln(NRq+1)/h_{\mathrm{p}}^*=0$.
\end{IEEEproof}

%------------------------------------------------------------------
\begin{rem} \label{rem:bound}
	The fact that in Appendix \ref{app:derivative} we require 
	$\rho_{\mathrm{s}}<1/2$ for \eqref{eq:der1} to be negative together with the 
	fact that we require
	$\rho_s=\rho_s'=\rho/(Rq\lambda)$ for $F_2$ to be zero
	results in the upper bound on the asymptotic minimum distance growth rate
	\begin{equation} \label{eq:UpperBoundDC}
			\hat{\rho}< \frac{Rq\lambda}{2}
	\end{equation}
	given in Definition~\ref{def:rhohat}.
\end{rem}
%------------------------------------------------------------------

We summarize our results in the following Theorem.
\begin{thm} \label{thm:main}
	If there exists a $\hat{\rho} > 0$ as defined in Definiton~\ref{def:rhohat} 
	for a 2-state TTC ensemble with $J>2$ 
	RCEs in the outer MPCC, then the 2-state TTC ensemble is asymptotically 
	good and the asymptotic minimum distance growth rate is at least $\hat{\rho}$.
\end{thm}
\begin{IEEEproof}
	From Lemmas~\ref{lem:a1} and \ref{lem:a2} we have that,
	for an arbitrarily small constant $\epsilon>0$, both
	\begin{equation*}
	\P{d_\mathrm{min} < \left\lceil N(\hat{\rho}-\epsilon) \right\rceil
						 \cap h_{\mathrm{p}} \leq h_{\mathrm{p}}^*}  \rightarrow 0
	\end{equation*}
	and
	\begin{equation*}
	\P{d_\mathrm{min} < \left\lceil N(\hat{\rho}-\epsilon) \right\rceil
						\cap h_{\mathrm{p}} > h_{\mathrm{p}}^*}  \rightarrow 0
	\end{equation*}
	as $N \rightarrow \infty$ for any $h_{\mathrm{p}}^*$ satisfying
	\begin{equation*}
		\lim_{N\rightarrow \infty} \frac{\ln (NRq+1)}{h^*_{\mathrm{p}}}=0 
		\quad \mathrm{and} \quad
		h^*_{\mathrm{p}} \leq N^{\frac{J-2}{J}-\epsilon}.
	\end{equation*}
\end{IEEEproof}
Theorem \ref{thm:main} proves that $\hat{\rho}$ is a lower bound on the 
asymptotic minimum distance growth rate of a 2-state TTC ensemble.
In a slight abuse of notation, from now on we refer to $\hat{\rho}$ as the 
asymptotic minimum distance growth rate.

%-----------------------------------------------------------------------------
\subsection{Asymptotic Minimum Distance Growth Rates}
\label{sub:numDist}

While the spectral shapes of TTCs cannot be negative, the existence of a 
positive $\hat{\rho}$ according to Definition~\ref{def:rhohat} implies that 
the ensemble is asymptotically good. As is common practice, we numerically 
evaluate the spectral shapes of TTC ensembles and use a subspace trust-region 
method \cite{Kelley99} to evaluate the supremum of the asymptotic CIOWE. 
For 2-state component codes, the asymptotic IOWDs are available in closed 
form, but for 4-state ensembles, we cannot obtain closed form expressions, 
so to compute the asymptotic spectral shapes we use the method outlined 
in \cite{STU02} to calculate them numerically.

\begin{figure}[t]
  \centerline{\includegraphics[width=1\columnwidth]{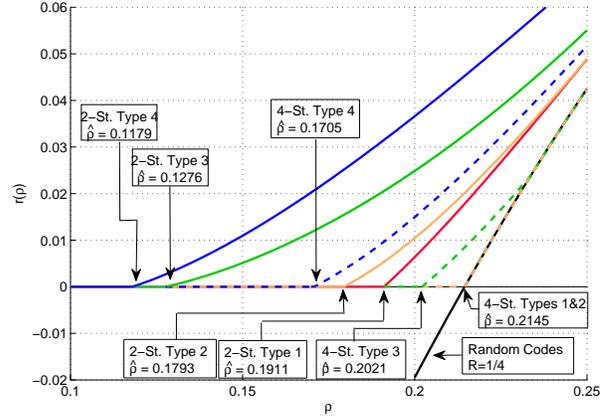}}
    \caption{Asymptotic spectral shapes for the rate $R=1/4$ 
    TTCs with $\mu=\lambda=1$.}
    \label{fig:SpShape}
\end{figure}

Fig.~\ref{fig:SpShape} shows the asymptotic spectral shapes for the
rate $R=1/4$ TTC ensembles with $\mu=\lambda=1$,
i.e., the spectral shapes of the HCCs.
The asymptotic spectral shape function of the entire ensemble of block codes
is also shown. It crosses zero at the GVB for rate $R=1/4$.
The ensembles with 2-state RCEs
in the outer MPCC are plotted  with solid lines, while the ensembles with
4-state RCEs in the outer MPCC are plotted with dashed lines.
The spectral shapes are never negative, but they
start out with a zero stretch and turn positive at the asymptotic distance 
growth rate $\hat{\rho}$.

Among the 2-state ensembles, the type 1 scheme has the largest asymptotic 
distance growth rate of $\hat{\rho}=0.1911$. 
Replacing one of the parallel concatenated accumulators by its feedforward 
inverse (type 2) decreases the asymptotic distance growth
rate to $\hat{\rho}=0.1793$. When only three branches enter the inner serially
concatenated accumulator and the output of the $1+D$ branch is sent straight
through to the channel (type 3), the asymptotic distance growth rate 
reduces further to $\hat{\rho}=0.1276$, and for the systematic type 4 scheme 
we obtain an asymptotic distance growth rate of only $\hat{\rho}=0.1179$.

Employing 4-state $[5/7]_8$ codes instead of accumulators in the outer MPCCs 
increases the asymptotic distance growth rates w.r.t. the 2-state ensembles.
In the case of the type 1 and type 2 ensembles with 4-state encoders
in the outer MPCCs, the positive part of the asymptotic spectral shape
is practically indistinguishable from the spectral shape of the
entire ensemble of block codes.

\begin{figure}[t]
    \centering
  \includegraphics[width=1\columnwidth]{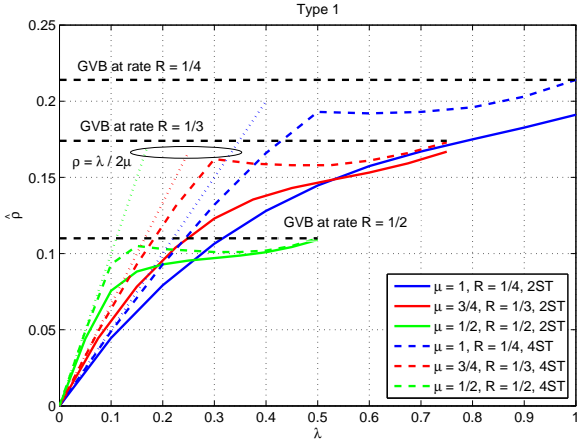}
    \caption{Asymptotic minimum distance growth rates of the type 1
                TTC as the tuning parameter
                $\lambda$ varies, $0\leq \lambda \leq \mu$. 
                In each case, the GVB is drawn only up to $\lambda = \mu$, 
                the maximum possible value of $\lambda$.}
    \label{fig:growthTP1}
\end{figure}
\begin{figure}[t]
    \centering
  \includegraphics[width=1\columnwidth]{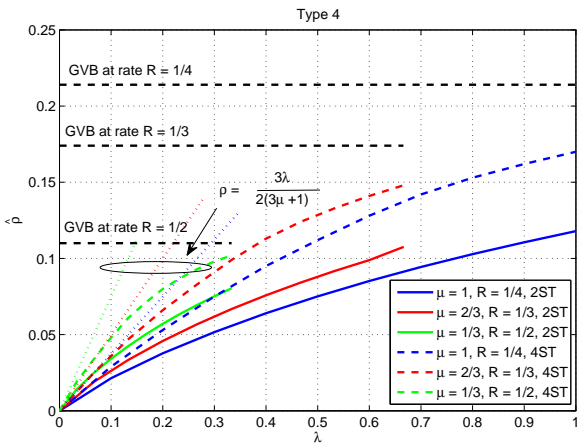}
    \caption{Asymptotic minimum distance growth rates of the type 4
                TTC as the tuning parameter
                $\lambda$ varies, $0\leq \lambda \leq \mu$.
                In each case, the GVB is drawn only up to $\lambda = \mu$, 
                the maximum possible value of $\lambda$.}
    \label{fig:growthTP4}
\end{figure}
Fig.~\ref{fig:growthTP1} shows the asymptotic minimum distance growth rates
$\hat{\rho}$ of the type 1 TTC as the tuning
parameter $\lambda$ varies, $0\leq \lambda \leq \mu$.
As the code rate increases by reducing the coefficient $\mu$, 
the initial slope of the asymptotic distance growth rate curve becomes steeper.
For small $\lambda$, these curves approach the upper bound on $\hat{\rho}$ given by
\eqref{eq:UpperBoundDC} and indicated by the line $\hat{\rho} = \frac{\lambda}{2\mu}$ 
for the type 1 ensemble.
This steep increase in the asymptotic distance growth rates with $\lambda$ is 
followed by the curve flattening out as the asymptotic distance growth rates approach 
the GVB. The green curve for the rate $R=1/2$ code ensemble shows the steepest 
increase (among the curves shown) but then flattens out around $\lambda = 0.15$. 
Like the 4-state code ensembles it reaches the 
GVB for $\lambda=\mu$. The asymptotic distance growth rates of the type 2 tuned 
turbo code show the same general behavior, but they are smaller than those 
of the type 1 ensemble and the increase for small $\lambda$ is not as steep.

Type 3 and 4 TTCs also show the same general behavior, with the asymptotic distance
growth rates of the type 3 ensemble being slightly larger than 
for the type 4 ensemble. The asymptotic distance growth rates of the type 4 ensemble 
are shown in Fig.~\ref{fig:growthTP4}. 
The initial slope of the curves is smaller than for the type 1 ensemble, and
for $\lambda=\mu$ the asymptotic distance growth rates are further away from the GVB. 
They also increase more smoothly with $\lambda$ than for the type 1 ensemble.

%-----------------------------------------------------------------------------
\subsection{Finite Length Analysis}
\label{sub:finDist}

The minimum distance of a TTC ensemble for a finite block length
$N$ can also be analyzed using \eqref{eq:dminBound}.
In particular, if we set $\P{\dmin<d} = \epsilon$, where $\epsilon$ is
any positive value between $0$ and $1$, we expect that at
least a fraction $1-\epsilon$ of the codes in the ensemble have a
minimum distance $\dmin$ of at least $d$. 
In the following, we choose $\epsilon = 1/2$, i.e., we expect that at least 
half of the codes in the ensemble have
a $\dmin$ at least equal to the value predicted by the curves.

\begin{figure}[t]
	\centering
		\includegraphics[width=1\columnwidth]{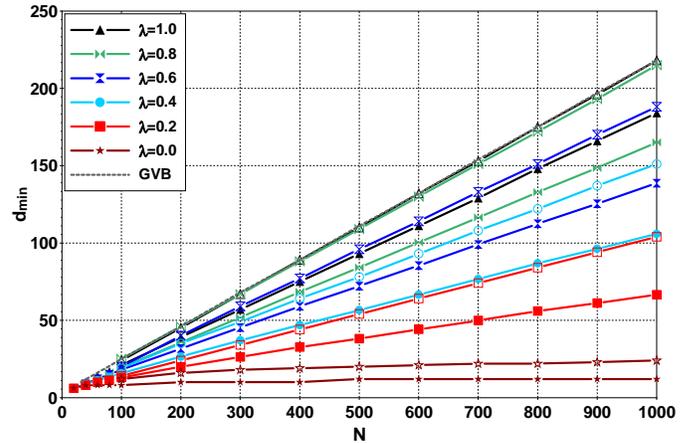}
		\caption{Lower bound on the minimum distance of the $R=1/4$ type 2 TTCs 
		with 2-state component encoders (filled markers) and 4-state component encoders 
		(empty markers) for $\mu=1$, $\epsilon=1/2$, and different values of the tuning 
		parameter $\lambda$.}
		\label{fig:distbounds}
\end{figure}
In Fig.~\ref{fig:distbounds} we show the lower bound on
$\dmin$ versus the code block length $N$ for the $R=1/4$ type 2 
tuned TTC ensembles with 2- and 4-state component 
encoders and several values of $\lambda$.
The finite length GVB is also plotted for reference.
The results are consistent with the asymptotic analysis in
the previous section and show increasing minimum distance growth
rates with increasing values of $\lambda$.
Also, for a given value of the tuning parameter
$\lambda$, the minimum distance of the type 2 code ensemble with
4-state component encoders is larger than for the code ensemble
with 2-state component encoders.

\begin{figure}[t]
	\centering
		\includegraphics[width=1\columnwidth]{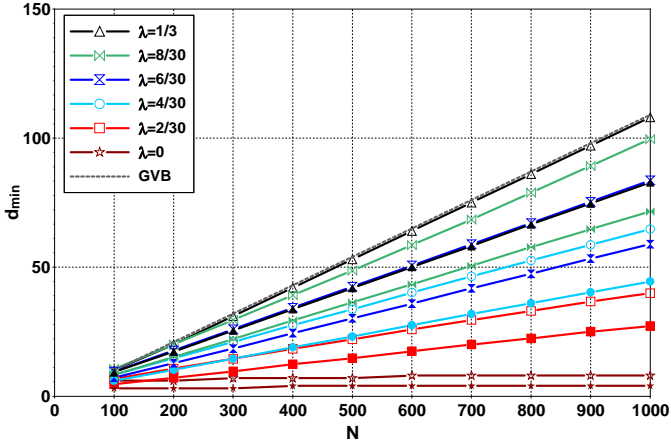}
		\caption{Lower bound on the minimum distance of the $R=1/2$ type 4 TTCs 
		with 2-state component encoders (filled markers) and 4-state component 
		encoders (empty markers) for $\mu=1/3$, $\epsilon=1/2$, and different values 
		of the tuning parameter $\lambda$.}
		\label{fig:distbounds4}
\end{figure}
In Fig.~\ref{fig:distbounds4} we display the same set of curves for the
$R=1/2$ type 4 TTCs with 2- and 4-state component encoders.
Again, the results are consistent with the asymptotic analysis, and
for a given value of the tuning parameter $\lambda$, the minimum distance 
of the code ensemble with 4-state component encoders is larger than for 
the code ensemble with 2-state component encoders.

For 4-state RCEs we cannot obtain a closed form WE.
However, since 2-state TTCs are asymptotically good, 
and replacing the accumulators in the outer MPCC
with more complex $[5/7]_8$ RCEs increases both the asymptotic 
distance growth rates (see Figs.~\ref{fig:SpShape}-\ref{fig:growthTP4}) 
as well as the finite block length minimum distances 
(see Figs.~\ref{fig:distbounds} and \ref{fig:distbounds4}), 
we strongly conjecture that the resulting code ensembles with the 
same structure are still asymptotically good.

%%%%%%%%%%%%%%%%%%%%%%%%%%%%%%%%%%%%%%%%%%%%%%%%%%%%%%%%%%%%%%%%%%%%%%%%%%%%%%
\section{Iterative Decoding Convergence Threshold} \label{sec:exit}

\begin{figure}[t]
  \centerline{\includegraphics[width=1\columnwidth]{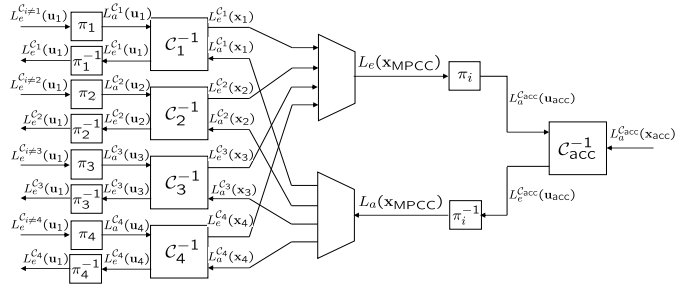}}
    \caption{Decoder for a type 1 TTC.}
    \label{fig:decoder}
\end{figure}

To determine the iterative decoding thresholds of tuned turbo code ensembles
we employ an extrinsic information transfer (EXIT) chart-based
analysis \cite{ten01TC}. EXIT charts track the exchange of extrinsic information 
between component decoders in a concatenated code scheme to estimate its iterative 
decoding threshold. In the following we briefly describe EXIT charts for type 1 TTCs. 
The decoder is depicted in Fig.~\ref{fig:decoder}. A similar
procedure as the one described below can also be applied to type 2, type 3,
and type 4 TTCs.

Let $\mathbf{u}_i=(u_0,\ldots,u_{K_{\mathcal{C}_i}-1})$ and 
$\mathbf{x}_i=(x_0,\ldots,x_{N_{\mathcal{C}_i}-1})$
be the sequence of information symbols and the sequence of code
symbols, respectively, of the $i$th $(N_{\calC_i},K_{\calC_i})$ component 
code $\calC_i$ of the TTC. (In the following, we will drop the index $i$ 
when referring to a generic component code.)  Each component decoder in 
Fig.~\ref{fig:decoder} is fed with \emph{a priori} information (from either 
other component decoders or the channel) on its information and coded symbols 
and computes extrinsic information which, in turn, is used by the other 
component decoders as \emph{a priori} information. In convergence analysis 
using EXIT charts it is common to model the \emph{a priori} information as 
a Gaussian random variable. Also, as required in EXIT charts analysis, we 
assume that $K_{\calC}\rightarrow\infty$ and $N_{\calC}\rightarrow\infty$.
For information symbol
$u$ the
corresponding \emph{a priori} L-value (or log-likelihood ratio) is denoted by
$L^\mathcal{C}_{\mathrm{a}}(u)$. Using the Gaussian approximation, 
$L^\mathcal{C}_{\mathrm{a}}(u)$ can be expressed as:
\begin{equation}
L^\mathcal{C}_{\mathrm{a}}(u)=\frac{\sigma_{\mathrm{a},u}^2}{2}u+w
\end{equation}
where $w$ is a zero-mean Gaussian random variable with variance 
$\sigma_{\mathrm{a},u}^2$. We denote by $I(u;L^\mathcal{C}_{\mathrm{a}}(u))$ the mutual
information (MI) between $u$ and $L^\mathcal{C}_{\mathrm{a}}(u)$. The average \emph{a
priori} MI for the information symbols is
\begin{equation}
    \begin{split}
        \IAuCode&=\frac{1}{K_\calC}\sum_{i=0}^{K_\calC-1}I(u_i;L^\mathcal{C}_{\mathrm{a}}(u_i)),
    \end{split}
\end{equation}
which depends only on $\sigma_{\mathrm{a},u}$ and can be computed using the $J$ 
function as $\IAuCode=\J{\sigma_{\mathrm{a},u}}$ \cite{ten01TC}.
Note that if $u$ is transmitted over the (binary-input Gaussian) channel, 
$L^\mathcal{C}_{\mathrm{a}}(u)$ corresponds to the channel L-value, 
$L_{\mathrm{ch}(u)}=4R\gamma r$, where $R$ is the code rate, $\gamma$ denotes 
the SNR $E_{\mathrm{b}}/N_0$, $r=\tilde{u}+n$ is the received observation, 
$\tilde{u}$ is the BPSK modulated symbol, and $n$ is AWGN with variance $N_0/2$. 
In this case it can be easily shown that $\sigma^2_{\mathrm{a},u}=8R\gamma$.

For code symbol $x$, the corresponding
\emph{a priori} L-value is denoted by $L^\mathcal{C}_{\mathrm{a}}(x)$. Using the 
Gaussian approximation $L^\mathcal{C}_{\mathrm{a}}(x)$ can be written as
\begin{equation}
L^\mathcal{C}_{\mathrm{a}}(x)=\frac{\sigma_{\mathrm{a},x}^2}{2}x+w,
\end{equation}
where $w$ is a zero-mean Gaussian random variable with variance $\sigma^2_{\mathrm{a},x}$. 
As before, we denote by $I(x;L^\mathcal{C}_{\mathrm{a}}(x))$ the MI
between $x$ and $L^\mathcal{C}_{\mathrm{a}}(x)$. The average \emph{a priori} MI for
the code symbols is given by
\begin{equation}
    \begin{split}
        \IAxCode=\frac{1}{N_\calC}\sum_{i=0}^{N_\calC-1}I(x_i;L^\mathcal{C}_{\mathrm{a}}(x_i)),
    \end{split}
\end{equation}
which can be computed using the $J$ function as $\IAxCode=\J{\sigma_{\mathrm{a},x}}$. 
If the code symbols are transmitted over the channel, $\sigma^2_{\mathrm{a},x}=8R\gamma$.

The \emph{a priori} L-values $L^\mathcal{C}_{\mathrm{a}}(u)$ and 
$L^\mathcal{C}_{\mathrm{a}}(x)$ are inputs to an
\emph{a posteriori} probability (APP) decoder which computes the extrinsic
L-values $L^\mathcal{C}_{\mathrm{e}}(u)$ and $L^\mathcal{C}_{\mathrm{e}}(x)$ 
for information symbols and code symbols, respectively. The extrinsic L-values 
are also Gaussian with variance $\sigma^2_{\mathrm{e},u}$ and $\sigma^2_{\mathrm{e},x}$, 
respectively. The average extrinsic MI for information
and code symbols is given by
\begin{equation}
    \IEuCode=\frac{1}{K_\calC}\sum_{i=0}^{K_\calC-1}I(u_i;L^\mathcal{C}_{\mathrm{e}}(u_i))
\end{equation}
and
\begin{equation}
    \IExCode=\frac{1}{N_\calC}\sum_{i=0}^{N_\calC-1}I(x_i;L^\mathcal{C}_{\mathrm{e}}(x_i)),
\end{equation}
respectively.

The input-output behavior of the APP decoder for encoder $\calC$ is
completely characterized by two EXIT functions, $T_u$ and
$T_x$, which specify the evolution of the extrinsic MIs as a
function of the \emph{a priori} MIs:
\begin{equation}
    \begin{split}
        \IEuCode&=T_u(\IAuCode,\IAxCode)\\
        \IExCode&=T_x(\IAuCode,\IAxCode).
    \end{split}
\end{equation}
In practice, these functions can be obtained by Monte Carlo
simulation for all values $0\leq\IAuCode\leq 1$ and
$0\leq\IAxCode\leq 1$ by modeling the \emph{a priori} information as
Gaussian distributed, as noted above. 

The decoder of the type 1 TTC consists of
$q+1$ APP component decoders
$\mathcal{C}^{-1}_1,\ldots,\mathcal{C}^{-1}_q$, and
$\mathcal{C}^{-1}_\mathrm{acc}$ corresponding to the component
encoders $\mathcal{C}_1,\ldots,\mathcal{C}_q$ of the outer MPCC and
to the inner accumulator, respectively, which iteratively exchange extrinsic 
information (see Fig.~\ref{fig:decoder}). A decoding iteration
consists of a single activation of
$\mathcal{C}^{-1}_1,\ldots,\mathcal{C}^{-1}_q$ and
$\mathcal{C}^{-1}_\mathrm{acc}$ in this order. The evolution of the
extrinsic MI can then be tracked in a multi-dimensional EXIT chart
\cite{Tue02}, which plots together the EXIT functions of the $q+1$ component 
encoders and can be used to predict the convergence
threshold. Unfortunately, such a multi-dimensional EXIT chart is
hard to visualize. To generate EXIT charts that are easier to deal
with, the EXIT functions of the component encoders of the outer MPCC
can be combined to obtain the EXIT function of the MPCC, without any precision 
loss in the prediction of the convergence thresholds \cite{Bra05IT}. In this way, 
the behavior of TTCs can be determined by using a two-dimensional EXIT chart, displaying 
in a single figure the EXIT functions of the outer MPCC and of the inner
recursive encoder:
\begin{equation}\label{eq:EXITfunctionsTTC}
    \begin{split}
        \IEyCodeMPCC & = T_x(\IAyCodeMPCC)\\
       \IExCodeff & = T_u(\IAxCodeff,\IAyCodeff),
    \end{split}
\end{equation}
where $\IAyCodeMPCC=\IExCodeff$ and $\IAxCodeff= \IEyCodeMPCC$. Note that, since the 
inner accumulator is connected to the channel, $\IAyCodeff$ is a function of $\gamma$. 
In particular, we must distinguish between the MI corresponding to the parity bits 
generated by the accumulator and the MI corresponding to the input bits, since the 
two branches are punctured with different puncturing rates. Assuming random puncturing, 
the \emph{a priori} MI for the parity
bits of the inner accumulator, punctured with rate
$\lambda$, is given by $\lambda J(\sqrt{8R\gamma})$, while the \emph{a priori} MI 
provided by the channel
for the input bits of the inner accumulator, punctured with
rate $\mu-\lambda$, is $(\mu-\lambda) J(\sqrt{8R\gamma})$. With these considerations, 
the EXIT function of the inner accumulator can be written as
\begin{equation}
\IExCodeff=T_u(\IEyCodeMPCC,(\lambda
J(\sqrt{8R\gamma}),(\mu-\lambda)
J(\sqrt{8R\gamma}))).
\end{equation}

What remains is the computation of $\IEyCodeMPCC$. Looking in more detail at the 
EXIT functions of the component encoders of the outer MPCC, we observe that the 
$l$th, $l=\{1,\hdots,4\}$, component decoder is fed with \emph{a priori} information 
on $\mathbf{u}_l$ generated by all the other component decoders of the MPCC, and 
with \emph{a priori} information on $\mathbf{x}_l$ provided by the decoder of the 
inner accumulator. The EXIT functions of the $l$th component of the outer MPCC can 
then be expressed as
\begin{equation}\label{eq:EXITcomp}
    \begin{split}
        \IExCodell &=T^{\mathcal{C}_l}_u\left(J\left(\sqrt{\sum_{i=1,i\neq
                l}^{L-1}J^{-1}\left(\IExCodeii\right)^2}\right),
        \IAyCodell\right)\\
        \IEyCodell &=T^{\mathcal{C}_l}_x\left(J\left(\sqrt{
                \sum_{i=1,i\neq l}^{L-1}J^{-1}\left(\IExCodeii\right)^2}\right),
        \IAyCodell\right),
    \end{split}
\end{equation}
for information symbols and code symbols, respectively. In (\ref{eq:EXITcomp})
we used the fact that 
$L^{\mathcal{C}_l}_{\mathrm{a}}(u)=\sum_{i\neq l}L^{\mathcal{C}_i}_{\mathrm{e}}(u)$ 
and (assuming independence) 
$\sigma^2_{\mathrm{a},u_l}=\sum_{i\neq l}\sigma^2_{\mathrm{e},u_i}$, 
from which (using the $J$ function) it follows that
$\IAxCodell=J\left(\sqrt{\sum_{i=1,i\neq
                l}^{L-1}J^{-1}\left(\IExCodeii\right)^2}\right)$ 
\cite{ten01TC}. Note that, for the type 1 tuned
turbo code, the four EXIT functions $\IExCodell$ and $\IEyCodell$ are identical and
$\IAyCodell=\IExCodeff$.
The EXIT function $\IEyCodeMPCC$ can be computed for all values
$0\leq\IExCodeff\leq 1$ by activating all $q$ decoders of the MPCC
until $\IExCodell$ and $\IEyCodell$ have converged to a fixed value. 
In other words, to obtain the two-dimensional EXIT plot, we assume that a 
large enough number of iterations is performed within the decoder of the outer 
MPCC before iterating with the decoder of the inner accumulator. Then, since 
all component encoders of the outer MPCC are identical, $\IEyCodeMPCC$ is just 
equal to $\IEyCodell$. Finally, the
convergence behavior of the type 1 TTC can be tracked by
displaying in a single plot the two EXIT functions in (\ref{eq:EXITfunctionsTTC}). 
The EXIT charts of type 2, type 3, and type 4 TTCs can be computed in a similar way. 
Note that  for type 3 and type 4, the EXIT function of the outer MPCC
also depends on $\gamma$ through encoder $\calC_0$, which is directly
connected to the channel. For the type 2 TTC, the
computation of $\IEyCodeMPCC$ is a bit more complex, since the EXIT
function of the first encoder in the outer MPCC is different.

\begin{figure}[t]
  \centerline{\includegraphics[width=1\columnwidth]{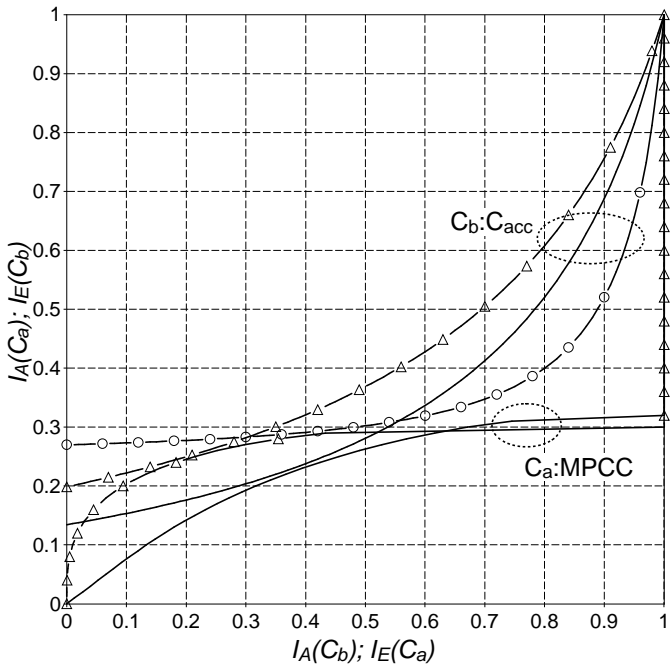}}
  \caption{EXIT charts of the type 4 TTC with $\lambda=1$ and
     $\gamma=1.03$ dB (solid curves with no markers), the type 1 TTC
    with $\lambda=1$ and $\gamma=2.24$ dB (triangles), and the type 1 TTC
    with $\lambda=0.3$ and
    $\gamma=1.09$ dB (circles). $R=1/4$, 2-state component
  encoders.}
    \label{fig:EXIT}
\end{figure}
In Fig.~\ref{fig:EXIT} we show the EXIT charts of the rate $R=1/4$
type 1 (triangles) and type 4 (solid curves with no markers) TTCs with
2-state component encoders for $\lambda=1$. A vertical step between
the lower curves and the upper curves represents a single activation
of the inner decoder, while a horizontal step between the upper
curves and the lower curves represents an unspecified number of
activations of all the component decoders of the MPCC until nothing
more can be gained. We observe that the type 4 TTC
converges significantly earlier ($\gamma=1.03$ dB) than the type 1
TTC ($\gamma=2.24$ dB), thanks to the systematic
branch. Note also that the EXIT chart for the type 1 tuned turbo
code is identical to that of the R$^4$AA code, where the EXIT
function of encoder $\mathcal{C}_{\mathrm{a}}$ in the figure corresponds now to
the EXIT function of the repeat-by-four code. The convergence
threshold of the type 1 TTC can be significantly
improved if some of the parity bits at the output of the inner
encoder are replaced by bits from the outer MPCC, at the expense of
a smaller asymptotic minimum distance growth rate. For the type 1 TTC with 
$\lambda=0.3$ (circles\footnote{Note that the EXIT function of the outer MPCC for 
type 1 TTCs is identical for $\lambda=1$ and $\lambda=0.3$, since it does not depend 
on $\lambda$.}) a tunnel opens at $\gamma=1.09$ dB, i.e., $1.15$ dB earlier.
In this case, the type 1 TTC with $\lambda=0.3$ has a
similar convergence threshold and asymptotic growth rate as the
type 4 TTC with $\lambda=1$.

%%%%%%%%%%%%%%%%%%%%%%%%%%%%%%%%%%%%%%%%%%%%%%%%%%%%%%%%%%%%%%%%%%%%%%%%%%%
\section{Tuning Behavior} \label{sec:Tuning}

In this section we combine the minimum distance results of Section \ref{sec:Dmin}
and the iterative decoding convergence results from Section \ref{sec:exit}.
We observed the tuning effect, namely asymptotic minimum distance growth rates
and iterative decoding thresholds increasing with $\lambda$, for all
types of tuned turbo codes. However, the effectiveness of tuning depends on 
the specific combination of distance and threshold results.

\begin{figure}[t]
    \centering
  		\includegraphics[width=1\columnwidth]{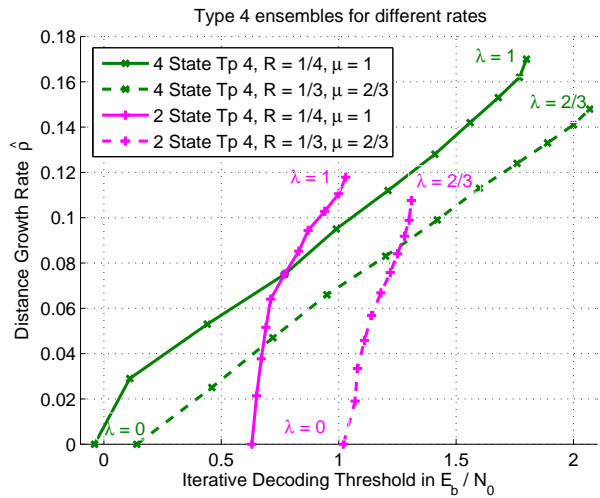}
    	\caption{Asymptotic minimum distance growth rate $\hat{\rho}$ 
    					versus the iterative decoding convergence threshold for the type 4
    					TTC with 2-state and 4-state encoders in the outer MPCC.}
    \label{fig:TuningTP4}
\end{figure}
Fig. \ref{fig:TuningTP4} shows the 
asymptotic minimum distance growth rate $\hat{\rho}$ 
versus the iterative decoding convergence threshold for the type 4 TTC 
with 2-state and 4-state encoders in the outer MPCC, respectively.
For all curves, we computed 11 equally spaced values from $\lambda=0$ to $\lambda=\mu$.
For $R=1/4$ ($\mu=1$) and $\lambda=1$, the ensemble with 4-state encoders
exhibits an asymptotic distance growth rate of $\hat{\rho}=0.17$ and a
threshold of $E_{\mathrm{b}}/N_0=1.8\,\mathrm{dB}$.  Decreasing $\lambda$ leads to
better convergence properties, but also to a reduction of the asymptotic distance growth rate.
In the extreme case of $\lambda=0$, the code is equal
to the outer MPCC consisting of a parallel concatenation of three
RCEs and a systematic branch. In this case, the minimum distance does not 
grow linearly with block length and the asymptotic distance growth rate 
therefore is zero.
Note that the outer MPCC with 4-state encoders has a 
significantly better iterative decoding convergence threshold
($E_{\mathrm{b}}/N_0=-0.04\,\mathrm{dB}$) than the MPCC with 2-state encoders
($E_{\mathrm{b}}/N_0=0.63\,\mathrm{dB}$).
For $\lambda=1$, the $R=1/4$ 2-state ensemble exhibits a better 
threshold ($E_{\mathrm{b}}/N_0=1.03\,\mathrm{dB}$) but a lower asymptotic distance growth 
rate ($\hat{\rho}=0.1179$) than the 4-state ensemble.
Therefore the dynamic range over which the 2-state ensemble can 
be adjusted is only 0.4\,dB, whereas the 4-state ensemble can be 
tuned over a larger range of thresholds and asymptotic distance growth rates.
This indicates that in the design of TTCs it is
important to use an outer MPCC with very good convergence properties.

Puncturing TTCs to rate $R=1/3$ ($\mu=2/3$) results in a right 
shift of the curves, while leaving their general shape intact. 
Since the maximum asymptotic distance growth rates (for $\lambda = \mu$) of the underlying 
$R=1/4$ code ensembles are not very close to the GVB, they are only 
slightly reduced by the puncturing process (see also Fig. \ref{fig:growthTP4}).

\begin{figure}[t]
    \centering
  		\includegraphics[width=1\columnwidth]{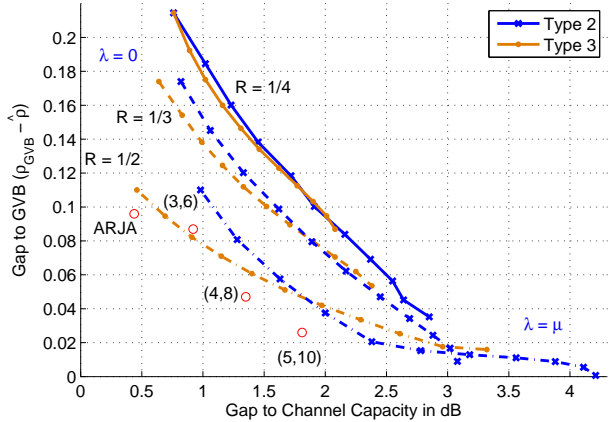}
    	\caption{Gap of the threshold to channel capacity versus the gap of 
    			the asymptotic minimum distance growth rate to the GVB for 2-state
    			type 2 and type 3 TTCs and rates $R=1/2$, $1/3$, and $1/4$. As a comparison, 
    			for $R=1/2$ the thresholds and asymptotic distance growth rate coefficients of some 
    			regular LDPC code ensembles, as well as the ARJA ensemble \cite{ADY07} are also given.}
    \label{fig:GapAbs23}
\end{figure}
In contrast to Fig. \ref{fig:TuningTP4}, which shows the values for 
the threshold and the asymptotic distance growth rate directly,  
in Fig. \ref{fig:GapAbs23} we show the gap between the convergence 
threshold and channel capacity and  
the gap between the asymptotic minimum distance growth rate and the GVB.
Since $\hat{\rho}=0$ for $\lambda=0$, the gap of the leftmost point of any curve is equal 
to the GVB. As $\lambda$ increases, the gap to the GVB decreases,
but the gap to channel capacity increases in all cases.
For $R=1/4$ and $\lambda=0$, the 2-state type 2 and type 3 ensembles are identical.
Due to the FFCE, they exhibit an iterative decoding threshold of
$E_{\mathrm{b}}/N_0=-0.04\,\mathrm{dB}$, only $0.75\,\mathrm{dB}$ from capacity.
For $\lambda>0$, the two ensembles exhibit somewhat different characteristics.

For $\mu=1$ $(R=1/4)$ and $\lambda=1$, the type 2 ensemble has an asymptotic distance growth rate 
of $\hat{\rho}=0.1793$, corresponding to a gap of $0.0352$ to the GVB,
and a threshold of $E_{\mathrm{b}}/N_0=2.05\,\mathrm{dB}$, corresponding to a gap to capacity 
of $2.85\,\mathrm{dB}$, while the type 3 ensemble exhibits a gap to the GVB of $0.0869$
and a gap to capacity of $2.08\,\mathrm{dB}$,
which is similar to the type 2 ensemble with $\lambda=0.6$.
While puncturing the code ensembles resulted in a right shift of the curves
in Fig. \ref{fig:TuningTP4}, in the representation of Fig. \ref{fig:GapAbs23}
puncturing moves the curves closer to the origin, i.e., for a fixed gap to capacity,
the gap to the GVB is smaller.
It is interesting to note that for $\lambda=0$ the gap to capacity of the type 2 ensemble
increases slightly as the rate increases, while for the type 3 ensemble
the gap to capacity decreases slightly as the rate increases.
The asymptotic distance growth rates for the $R=1/2$ type 2 TTC 
behave like those shown in Fig. \ref{fig:growthTP1} for type 1 ensembles. 
For small values of $\lambda$ they rapidly increase and then flatten out as 
the asymptotic distance growth rate approaches the GVB. The iterative decoding 
threshold, however, continuously increases with $\lambda$, so that
the tuning behavior of the $R=1/2$ type 2 ensemble flattens for a stretch 
before it reaches the GVB at $\lambda =\mu = 1/2$. Therefore the parameter range 
over which the ensemble can be effectively tuned is from $\lambda=0$ to 
$\lambda = 0.2$, which brings the asymptotic distance growth rate to within 0.02 of the GVB.

As a comparison we also give the threshold and asymptotic distance growth rates for rate 
$R=1/2$ regular LDPC code ensembles and the rate $R=1/2$ ARJA \cite{ADY07} ensemble.
With the exemption of the (3,6) LDPC code ensemble, for a given gap to channel capacity, 
the LDPC code ensembles exhibit a larger asymptotic distance growth rate than the TTC ensembles.
However, in contrast to the asymptotically good LDPC codes, TTCs have a simple encoder structure
with $O(1)$ encoding complexity. 
On the other hand, the quasi-cyclic subensemble of the above LDPC codes that also has
$O(1)$ encoding complexity is not asymptotically good.

\begin{figure}[t]
    \centering
  		\includegraphics[width=1\columnwidth]{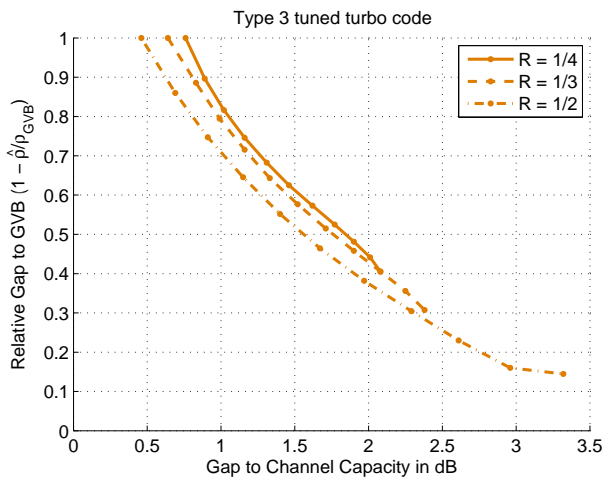}
    	\caption{Gap of the threshold to channel capacity versus the relative gap of 
    			the asymptotic minimum distance growth rate to the GVB for the 2-state
    			type 3 TTC ensemble with rates $R=1/2$, $1/3$, and $1/4$.}
    \label{fig:GapRel3}
\end{figure}
Fig. \ref{fig:GapRel3} again shows the tuning behavior of the 2-state type 3 ensemble,
but this time the y-axis shows the relative distance from the GVB, namely 
$1-\frac{\hat{\rho}}{\rho_{GVB}}$. 
The slopes of the three curves are almost identical.
Therefore improving the asymptotic distance growth rate from zero to half of the GVB in 
each case corresponds to a difference in the convergence threshold of roughly 1~dB.
The curves plot the maximum possible range of $\lambda$ values, with $\lambda=0$ 
corresponding to the topmost point and $\lambda=\mu$ corresponding to the lowest point of each curve.

%%%%%%%%%%%%%%%%%%%%%%%%%%%%%%%%%%%%%%%%%%%%%%%%%%%%%%%%%%%%%%%%%%%%%%%%%%%
\section{Simulation Results} \label{sec:Sim}

While the previous sections focused on asymptotic results for the minimum 
distance and the iterative decoding convergence behavior, in this section we show
simulation results illustrating that the tuning principle also applies 
to relatively short block lengths. 
\begin{figure}[t]
  \centerline{\includegraphics[width=1\columnwidth]{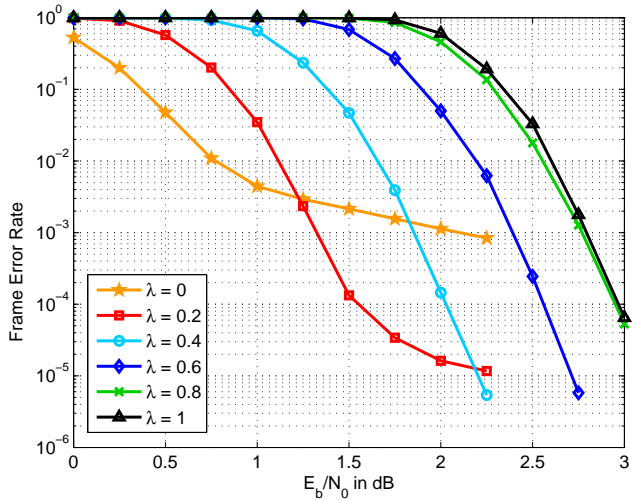}}
	\caption{Frame error rate performance of $R=1/4$ ($\mu=1$) type 2 TTCs with 2-state 
	component encoders for different values of the tuning parameter $\lambda$.}
    \label{fig:simulations}
\end{figure}
We did not make any attempt to optimize the 
simulated codes but rather focused on the ensemble average code performance.
To this end, random interleavers, as well as random puncturing patterns, were employed. 
Carefully designing the interleavers and puncturing patterns should yield better codes 
than the ones shown here in terms of error floor performance \cite{HLCZ06}. Interleaver design,
however, usually has little influence on the iterative convergence threshold.
The information block length for all simulations is $K=1024$ bits
and we use 20 iterations.

\begin{figure}[t]
  \centerline{\includegraphics[width=1\columnwidth]{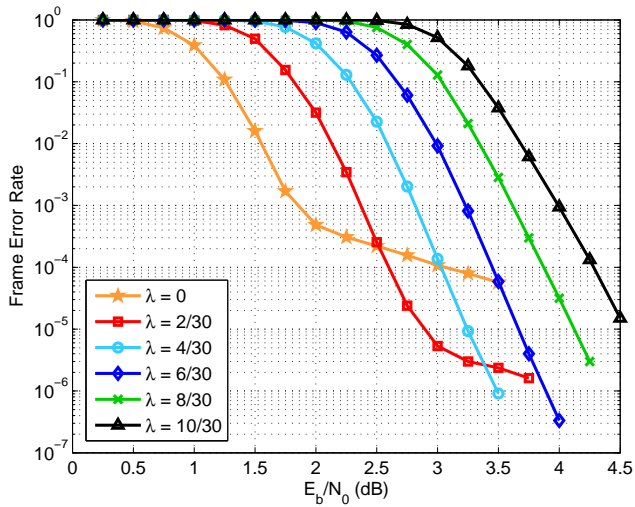}}
    \caption{Frame error rate performance of R=1/2 type 4 TTCs with 4-state
     component encoders for different values of the tuning parameter $\lambda$.}
    \label{fig:sim4states}
\end{figure}

In Fig.~\ref{fig:simulations}, we display frame error rate (FER) curves for rate $R=1/4$ 
type 2 tuned turbo codes with 2-state component encoders and
$\lambda\in[0,1]$.
The type 2 code with $\lambda=0$ performs best in the
waterfall region, but it has a high error floor due to its poor
minimum distance\footnote{The height of the error floor of the MPCCs ($\lambda=0$) in 
Figs.~\ref{fig:simulations} and \ref{fig:sim4states} is accurately predicted by the 
union bound of the code, indicating that the dominant cause of decoding error is decoding 
to a wrong codeword. For $\lambda>0$, the error floor is above the union bound, indicating 
that the minimum pseudo-weight of the code is limiting performance in the error floor region.}.
In this case, the code is equivalent to the MPCC in \cite{MaC01}. 
On the other hand, the code with $\lambda = 1$ shows the worst convergence, but according 
to the analysis in Section~\ref{sec:Dmin}, it has the best asymptotic minimum distance growth 
rate, potentially resulting in the lowest error floor.
By tuning $\lambda$, we can obtain any behavior in
between these two extreme cases: when $\lambda$ decreases, the
convergence behavior of the code improves (the curves get closer to the performance 
of the MPCC), but the error floor is higher.
For small values of $\lambda$, where the minimum distance is small, 
the simulations were able to reach the error floor of the code.
Compared to $\lambda=0$, the code with $\lambda=0.2$ loses about 0.5~dB 
in the waterfall region but the height of the error floor improves
by two orders of magnitude. For $\lambda=0.4$, the convergence threshold is again
0.5~dB worse than the $\lambda=0.2$ case,
but the error floor is lowered beyond what can be observed in the simulations.

Similar behavior is observed in Fig.~\ref{fig:sim4states}, 
where FER curves for rate $R=1/2$ type 4 TTCs 
with 4-state component encoders are shown.
Again, by varying the tuning parameter $\lambda$, we can obtain any
behavior between the outer MPCC, which shows the best iterative decoding
convergence behavior, and the HCC, which has the best error floor. 
In particular, the code with $\lambda=0$ performs best in the waterfall region,
as predicted by the EXIT charts. However, it has the highest error
floor, in agreement with the minimum distance analysis.
In general, lower error floors are obtained for increasing values of $\lambda$,
but at the expense of poorer performance in the waterfall region.
Note that, due to the more powerful 4-state component encoders employed, the error 
floors of the $R=1/2$ type 4 TTCs are lower than those observed in 
Fig.~\ref{fig:simulations} for the $R=1/4$ type 2 TTCs with 
2-state component encoders.

%%%%%%%%%%%%%%%%%%%%%%%%%%%%%%%%%%%%%%%%%%%%%%%%%%%%%%%%%%%%%%%%%%%%%
\section{Conclusions} \label{sec:Concl}

In this paper, we have introduced a family of hybrid concatenated codes
where a tradeoff between asymptotic minimum distance growth rate and iterative decoding
threshold can be achieved by varying a tuning parameter $\lambda$. By
decreasing $\lambda$, the convergence behavior of the code is improved
at the expense of a smaller asymptotic minimum distance growth rate and worse error floor
performance, and vice versa.
An important advantage of the hybrid tuned turbo code constructions is
that they are asymptotically good for a large range of values of $\lambda$,
so that even small values of $\lambda$ are sufficient to ensure linear asymptotic
distance growth with block length, potentially resulting in low error floors.
In addition, a second tuning parameter $\mu$ can be used to change the rate of
a TTC ensemble, thereby allowing a system designer to trade off between
code rate, iterative decoding convergence behavior, and error floor performance
without changing the encoder structure.

%%%%%%%%%%%%%%%%%%%%%%%%%%%%%%%%%%%%%%%%%%%%%%%%%%%%%%%%%%%%%%%%%%%%%
\appendices

\section{Proof of Proposition \ref{prop:concave}} 
\label{app:concave}

A sequence $\alpha_i$, $i=0,1,\ldots,n$, is logarithmically concave if
    \begin{equation*}
        \alpha_i^2 \geq \alpha_{i-1} \cdot \alpha_{i+1}
    \end{equation*}
    holds for every element $\alpha_i$ with $1\leq i \leq n-1$ \cite{S89}.
    
    \begin{enumerate}
    \item For the accumulator, we now consider the ratio
    \begin{equation*}
      \begin{aligned}
        R_w = & \frac{\left(\mathbb{P}^{\mathrm{Acc}(N_\calC)}_{w,h}\right)^2}
        {\mathbb{P}^{\mathrm{Acc}(N_\calC)}_{w,h-1} \mathbb{P}^{\mathrm{Acc}(N_\calC)}_{w,h+1}}\\
          = & \frac{N_\calC-h}{N_\calC-h+1} \cdot
        \frac{N_\calC-h-\left\lfloor w/2 \right\rfloor +1}
        	{N_\calC-h-\left\lfloor w/2 \right\rfloor} \cdot\\
        & \frac{h-1}{h} \cdot
        \frac{h-\left\lceil w/2 \right\rceil +1}{h-\left\lceil w/2 \right\rceil}.
      \end{aligned}
    \end{equation*}
    Since the ratio $\frac{x_1}{x_1+1}\cdot \frac{x_2+1}{x_2} > 1$ for $x_1>x_2$, we
    obtain $R_1 = 1$ for $w=1$ and $R_w>1$ for $w>1$.
    The sequence is thus logarithmically concave.
    Since the logarithm is a monotonically increasing function, the maximum of the
    IOWD equals the maximum of the asymptotic IOWD and can thus
    be obtained by taking the derivative of \eqref{eq:AccAIOWD}, which is given by
    \begin{equation} \label{eq:diffPacc}
            \frac{\partial}{\partial \beta} f^{\mathrm{Acc}}_{\alpha,\beta}
                    = \ln \left( \frac{\beta}{1-\beta} \right)
                        + \ln \left( \frac{1-\beta-\alpha/2}{\beta - \alpha/2} \right),
    \end{equation}
    so the maximum occurs at $\beta=1/2$, where $f^{\mathrm{Acc}}_{\alpha, 1/2}=0$.
    Correspondingly, the IOWD of the accumulator is maximized for $h=N_\calC/2$.

    \item For simplicity, we consider the terminated 2-state FFCE with even output weight $h$.
    (Considering the terminated code does not change the asymptotic IOWD of the code.)
    The ratio $R_w$ is given by
    \begin{equation*}
     \begin{aligned}
        R_w = & \frac{\left(\mathbb{P}^{{\mathrm{FF}}(N_\calC)}_{w,h}\right)^2}
        {\mathbb{P}^{{\mathrm{FF}}(N_\calC)}_{w,h-2} \mathbb{P}^{{\mathrm{FF}}(N_\calC)}_{w,h+2}}\\
          = & \frac{h/2+1}{h/2-1} \cdot
        \frac{N_\calC-w-h/2+1}{N_\calC-w-h/2-1}\cdot
        \frac{w-h/2 +1}{w-h/2 -1} >1.
     \end{aligned}
    \end{equation*}
    The sequence is strictly logarithmically concave since every term in the above product 
    is strictly larger than one.
    The derivative of \eqref{eq:1pDAIOWD} is given by
    \begin{equation} \label{eq:diffP1pD}
            \frac{\partial}{\partial \beta} f^{\mathrm{FF}}_{\alpha,\beta}
                    = \frac{1}{2} \ln \left( \frac{\alpha-\beta/2}{\beta/2} \right)
                    + \frac{1}{2} \ln \left( \frac{1-\alpha-\beta/2}{\beta/2} \right),
    \end{equation}
    so the maximum occurs at $\beta=2\alpha(1-\alpha)$, where 
    $f^{\mathrm{Acc}}_{\alpha, 2\alpha(1-\alpha)}=0$.
    Correspondingly, the IOWD of the 2-state FFCE is maximized for 
    $h=2w( 1-w/{N_\calC})$.

\item For random puncturing, we consider the ratio
\begin{equation*}
     \begin{aligned}
        R_h = & \frac{\left(\mathbb{P}^{\mathrm{P}(N')}_{h,h',\lambda}\right)^2}
        {\mathbb{P}^{\mathrm{P}(N')}_{h,h'-1,\lambda} \mathbb{P}^{\mathrm{P}(N')}_{h,h'+1,\lambda} }\\
          = & \frac{h'+1}{h'} \cdot
                \frac{\lambda N-h'+1}{\lambda N-h'} \cdot\\
        & \frac{h-h'+1}{h-h'} \cdot
        \frac{(1-\lambda )N-h+h'+1}{(1-\lambda )N-h+h'} > 1.
     \end{aligned}
\end{equation*}
The sequence is strictly logarithmically concave since every term in the above product is strictly larger than one.
The derivative of \eqref{eq:PunctAIOWD} is given by
\begin{equation*} \label{eq:diffPunct}
            \frac{\partial}{\partial \beta'} f^{\mathrm{P}}_{\beta,\beta',\lambda}
                    = -\ln \left( \frac{\beta'}{1-\beta'} \right)
                        +\ln \left( \frac{\beta-\lambda \beta'}
                                    {1-\lambda -\beta +\lambda\beta'} \right),
\end{equation*}
    so the maximum occurs at $\beta'=\beta$, where $f^{\mathrm{P}}_{\beta, \beta, \lambda}=0$.
 Correspondingly, the IOWD of the random puncturing operation is maximized for $h'=\lambda h$.
\end{enumerate}
 
%%%%%%%%%%%%%%%%%%%%%%%%%%%%%%%%%%%%%%%%%%%%%%%%%%%%%%%%%%%%%%%%%%%%%%%%%
\section{The Total derivative for $F_1$}
\label{app:derivative}

We bound the total derivative of 
\begin{equation*}
	F_1 = \frac{1}{q}\H{ \alpha } 
         		+ \frac{\calI_0}{q} f^{\calC_0}_{\alpha, \rho_0} 
            + \frac{J}{q} f^{\mathrm{Acc}}_{\alpha,\rho_\mathrm{r}}
            + \frac{q-J}{q} f^{\mathrm{FF}}_{\alpha,\rho_{\mathrm{f}}}
            + f^{\mathrm{Acc}}_{\rho_{\mathrm{p}},\, \rho_{\mathrm{s}}}
\end{equation*}
as $\rho_{\mathrm{p}} \rightarrow 0$.

To capture the dependency of $\rho_{\mathrm{p}}$ on the normalized input weight $\alpha$, 
the normalized output weight of the 2-state FFCEs $\rho_{\mathrm{f}}$ and the normalized 
weight of the systematic branch $\rho_0$, we parameterize $\alpha$ as 
$\alpha = a \rho_{\mathrm{r}}$, with $a \in [0,2]$, where the range of $a$ follows from 
the fact that the output weight of the accumulator cannot be less than half the input weight.
Likewise we parameterize $\rho_{\mathrm{f}} = b\alpha = ab\rho_{\mathrm{r}}$, with $b \in [0,2]$.
With the above parameterization, the weights $\rho_{\mathrm{p}}$ and $\rho_{\mathrm{r}}$ are 
related by a multiplicative factor, i.e., $\rho_{\mathrm{p}} = \rho_{\mathrm{r}}(J+ab(q-J))/q$.

We parameterize the output weight of the optional parallel encoder as 
$\rho_0 = c\alpha = ac\rho_{\mathrm{r}}$, where $c=0$ if there is no parallel encoder, $c=1$ if there 
is a simple systematic branch, and $c \in [0,2]$ if the parallel encoder is the 2-state FFCE.

The total derivative w.r.t. $\rho_{\mathrm{p}}$ is then given by
\begin{equation*} \label{eq:totalDerF1}
    \begin{aligned}
        \frac{d}{d \rho_p} & \left. F_1 
        			\right|_{\substack{\alpha= a\rho_{\mathrm{r}} \\
        		  \rho_{\mathrm{f}}= ab\rho_{\mathrm{r}} \\ \rho_0 = ac \rho_{\mathrm{r}}}} 			
       = \frac{\partial}{\partial \rho_{\mathrm{p}}} F_1 + \\
       		& \frac{q}{J+ab(q-J)} \left( \frac{\partial}{\partial \rho_{\mathrm{r}}} F_1
            + \frac{\partial}{\partial \alpha} F_1
              \cdot \frac{\partial \alpha}{\partial \rho_{\mathrm{r}}}+\right.\\        
          & \left.\left. \frac{\partial}{\partial \rho_{\mathrm{f}}} F_1 
              \cdot \frac{\partial \rho_{\mathrm{f}}}{\partial \rho_{\mathrm{r}}}
      		  + \frac{\partial}{\partial \rho_0} F_1
              \cdot \frac{\partial \rho_0}{\partial \rho_{\mathrm{r}}}
              \right) \right|_{\substack{\alpha= a\rho_{\mathrm{r}} 
        			\\ \rho_{\mathrm{f}}= ab\rho_{\mathrm{r}} \\ \rho_0 = ac \rho_{\mathrm{r}}}}.
     \end{aligned}
\end{equation*}

The derivatives of the asymptotic IOWDs of the component encoders with respect to their 
output weight are given by \eqref{eq:diffPacc} and \eqref{eq:diffP1pD}. The derivatives 
with respect to their input weight are given by
\begin{equation} \label{eq:diffPacc1}
\begin{aligned}
	\frac{\partial}{\partial \alpha} f^{\mathrm{Acc}}_{\alpha,\beta} = &
          \frac{1}{2}\ln \left( \frac{\beta - \alpha/2}{1-\alpha} \right)
             + \frac{1}{2}\ln \left( \frac{1-\beta - \alpha/2}{1-\alpha} \right) 
             + \ln 2 \\
           = & \frac{1}{2}\ln \left( 4x(1-x) \right),
\end{aligned}
\end{equation}
where $x=\frac{\beta-\alpha/2}{1-\beta}$, and
\begin{equation} \label{eq:diffP1pD1}
            \frac{\partial}{\partial \alpha} f^{\mathrm{FF}}_{\alpha,\beta}
                    = 2 \ln \left( \frac{\alpha}{1-\alpha} \right)
                        + \ln \left( \frac{1-\alpha-\beta/2}{\alpha - \beta/2} \right).
\end{equation}

We now evaluate the contributions of the component encoders in $F_1$ to the total 
derivative and show that for $\rho_\mathrm{p} \rightarrow 0$, or equivalently 
$\rho_\mathrm{r} \rightarrow 0$, the contribution of each component is either zero 
or negative. For convenience, for the component encoders of the outer MPCC and 
the systematic branch we consider the derivative w.r.t $\rho_{\mathrm{r}}$ rather 
than the derivative w.r.t. $\rho_{\mathrm{p}}$.
When the parallel encoder is simply a systematic branch, its contribution to the total 
derivative is a constant.
When the parallel encoder is the 2-state FFCE, its contribution to 
the total derivative is given by

\begin{equation*} \label{eq:diffPart1}
    \begin{aligned}
           \frac{\calI_0}{q} \frac{\partial}{\partial \alpha} & 
               		\left. f^{\mathrm{FF}}_{\alpha,\rho_0}\right|_{\substack{\alpha= a\rho_{\mathrm{r}} 
        			\\ \rho_0 = ac\rho_{\mathrm{r}} }} \frac{\partial \alpha}{\partial \rho_\mathrm{r}} +
        			\frac{\calI_0}{q} \frac{\partial}{\partial \rho_0}
        				\left. {f^{\mathrm{FF}}_{\alpha,\rho_0}}\right|_{\substack{\alpha= a\rho_{\mathrm{r}} 
        			\\ \rho_0 = ac \rho_{\mathrm{r}}}} \frac{\partial \rho_0}{\partial \rho_\mathrm{r}}  \\
        			%%%
        			= & \frac{\calI_0a}{q} \left[ 
        			\ln \left( \frac{1-a\rho_{\mathrm{r}}-ac\rho_{\mathrm{r}} /2}{1 - a\rho_{\mathrm{r}}}\right)
        			- \ln\left(1-\frac{c}{2}\right)+\right.\\
        		& \left. \ln\left( \frac{a\rho_{\mathrm{r}}}{1-a\rho_{\mathrm{r}}} \right)\right] +
        		  \frac{\calI_0ac}{2q} \left[ 
        		  \ln \left( \frac{1-a\rho_{\mathrm{r}}-ac\rho_{\mathrm{r}}/2}{1-a\rho_{\mathrm{r}}}\right)+\right.\\
        		& \left. \ln\left( \frac{a\rho_{\mathrm{r}}(1-c/2)}{a\rho_{\mathrm{r}}} \right)
        		  - \ln\left( \frac{ac\rho_{\mathrm{r}}/2}{1-a\rho_{\mathrm{r}}} \right)
        			- \ln\left( \frac{ac\rho_{\mathrm{r}}/2}{a\rho_{\mathrm{r}}}\right) \right] \\
        			%%%
        		= & \frac{\calI_0a}{q} \left[ 
        			\underbrace{ \ln \left( \frac{(c/2)^{-c/2}(a\rho_{\mathrm{r}})^{1-c/2}}
        			{(1-a\rho_{\mathrm{r}})^{1-c/2}} \right) }_1 + \right.\\
        		&	\underbrace{ \left(\frac{c}{2}+1\right) 
        			\ln \left( \frac{1-a\rho_{\mathrm{r}}(1+c/2)}{1 - a\rho_{\mathrm{r}}}\right)}_2+\\
        		& \left. \underbrace{ \left(\frac{c}{2}-1\right)\ln\left(1-\frac{c}{2}\right) }_3
        			- \underbrace{ \frac{c}{2}\ln\left(\frac{c}{2}\right) }_4 \right].
    \end{aligned}
\end{equation*}
For $a=0$, the above expression takes on the value zero. 
For any fixed $a$, $0 < a \leq 2$, term 1 is zero for $c=2$ and tends to $-\infty$ otherwise. 
Term 2 is zero for $c=0$ and strictly negative otherwise, and it vanishes as 
$\rho_\mathrm{r} \rightarrow 0$.
Terms 3 and 4 are constants and are zero for $c=0$ and $c=2$.

Similarly, the contribution of the FFCEs in the outer MPCC to the total derivative is given by
\begin{equation*} \label{eq:diffFFCEs}
    \begin{aligned}
           \frac{q-J}{q} \frac{\partial}{\partial \alpha} & 
               		\left. f^{\mathrm{FF}}_{\alpha,\rho_\mathrm{f}}\right|_{\substack{\alpha= a\rho_{\mathrm{r}} 
        			\\ \rho_\mathrm{f} = ab\rho_{\mathrm{r}} }} \frac{\partial \alpha}{\partial \rho_\mathrm{r}} +
        			\frac{q-J}{q} \frac{\partial}{\partial \rho_\mathrm{f}}
        				\left. {f^{\mathrm{FF}}_{\alpha,\rho_\mathrm{f}}}\right|_{\substack{\alpha= a\rho_{\mathrm{r}} 
        			\\ \rho_\mathrm{f} = ab \rho_{\mathrm{r}}}} \frac{\partial \rho_\mathrm{f}}
        			{\partial \rho_\mathrm{r}}  \\
        			%%%
        			= & \frac{a(q-J)}{q} \left[ 
        			\ln \left( \frac{(b/2)^{-b/2}(a\rho_{\mathrm{r}})^{1-b/2}}
        			{(1-a\rho_{\mathrm{r}})^{1-b/2}} \right) + \right.\\
        		& \left(\frac{b}{2}+1\right) 
        			\ln \left( \frac{1-a\rho_{\mathrm{r}}(1+b/2)}{1 - a\rho_{\mathrm{r}}}\right)+\\
        		& \left. \left(\frac{b}{2}-1\right)\ln\left(1-\frac{b}{2}\right)
        			- \frac{b}{2}\ln\left(\frac{b}{2}\right) \right],
    \end{aligned}
\end{equation*}
and either tends to zero or $-\infty$ as $\rho_\mathrm{r} \rightarrow 0$.

The contribution of the RCEs in the outer MPCC to the total derivative is given by
\begin{equation*} \label{eq:diffRCEs}
    \begin{aligned}
           \frac{\partial}{\partial \alpha} & 
               		\left.\left( \frac{1}{q}\H{\alpha} + \frac{J}{q} f^{\mathrm{Acc}}_{\alpha,\rho_\mathrm{r}}
               		\right) \right|_{\substack{\alpha= a\rho_{\mathrm{r}}}} 
              \frac{\partial \alpha}{\partial \rho_\mathrm{r}} +
        			\frac{\partial}{\partial \rho_\mathrm{r}} \frac{J}{q}
        				\left. {f^{\mathrm{Acc}}_{\alpha,\rho_\mathrm{r}}}\right|_{\substack{\alpha= a\rho_{\mathrm{r}}}} \\
							%%%
        		= & \frac{aJ}{q} \left[ -\frac{1}{J} \ln\left(
        		    \frac{a\rho_\mathrm{r}}{1-a\rho_\mathrm{r}} \right)
        			+ \frac{1}{2}\ln \left( \frac{\rho_{\mathrm{r}}(1-a/2)} {1-a\rho_{\mathrm{r}}} \right)+\right.\\
            & \left. \frac{1}{2}\ln \left( \frac{1-\rho_{\mathrm{r}} 
             		- a\rho_{\mathrm{r}}/2}{1-a\rho_{\mathrm{r}}} \right)
              + \ln 2 \right]+ \\
            &   \frac{J}{q} \left[ \ln\left( \frac{\rho_{\mathrm{r}}}{1-\rho_{\mathrm{r}}} \right)
              + \ln \left( \frac{1-\rho_{\mathrm{r}}-a\rho_{\mathrm{r}} /2}
             			{\rho_{\mathrm{r}} - a\rho_{\mathrm{r}}/2} \right) \right] \\
             %%%%
        		= &  \frac{J}{q} \left[ \underbrace{ \frac{a}{2} \ln \left( \frac{4(a\rho_{\mathrm{r}})^{1-2/J}}
        				{a(1-a\rho_{\mathrm{r}})^{1-2/J}} \right) }_1
        		 + \underbrace{  \frac{a}{2} \ln \left( \frac{1-\rho_{\mathrm{r}}(1+a/2)} 
        		 			{1-a\rho_{\mathrm{r}}} \right) }_2+\right. \\
    	& \left. \underbrace{ \ln \left( \frac{1-\rho_{\mathrm{r}}(1+a/2)} 
        		 			{1-\rho_{\mathrm{r}}} \right) }_3
        		 + \underbrace{ \left( \frac{a}{2}-1 \right) \ln\left( 1-\frac{a}{2}\right) }_4 \right].
    \end{aligned}
\end{equation*}
For $a=0$, the above expression is zero. For any fixed $a$, $0 < a \leq 2$, and $J>2$, 
term 1 tends to $-\infty$ as $\rho_\mathrm{r} \rightarrow 0$.
Terms 2 and 3 vanish for $\rho_{\mathrm{r}} \rightarrow 0$ and term 4 is a constant.

Finally, the contribution of the inner accumulator to the total derivative is given by
\begin{equation} \label{eq:diffIAcc}
            \frac{\partial}{\partial \rho_{\mathrm{p}}} 
            f^{\mathrm{Acc}}_{\rho_{\mathrm{p}}, \rho_{\mathrm{s}}}
           = \frac{1}{2}\ln \left( 4x(1-x) \right),
\end{equation}
with $x=\frac{\rho_{\mathrm{s}}-\rho_{\mathrm{p}}/2}{1-\rho_{\mathrm{p}}}$, which is negative for all
$x<1/2$, or equivalently $\rho_{\mathrm{s}}<1/2$.

Thus, using $\rho_s < \hat{\rho}/(Rq\lambda)$ and the fact that \eqref{eq:diffIAcc} 
is concave in $x$, for any $0 \leq \rho < \hat{\rho} < Rq\lambda/2$, we have 
\begin{equation*} \label{eq:TotalDer1c}
        \frac{d}{d \rho_{\mathrm{p}}} \left. F_1 \right|_{\rho_{\mathrm{p}}=0} \leq
        		\frac{1}{2}\ln \left( 4\frac{\hat{\rho}}{Rq\lambda}
        		\left(1-\frac{\hat{\rho}}{Rq\lambda}\right) \right) = -C_2,
\end{equation*}
where we have used the fact that, for $\rho_{\mathrm{p}} \rightarrow 0$, $x \rightarrow \rho_{\mathrm{s}}$.

%----------------------------------------------------------------------

\section*{Acknowledgment}
The authors would like to thank the associate editor for his thorough reading of 
the manuscript and for his comments that greatly improved the presentation of the paper.

%%%%%%%%%%%%%%%%%%%%%%%%%%%%%%%%%%%%%%%%%%%%%%%%%%%%%%%%%%%%%%%%%%%%%%%%%%%%%%%%%%
%\bibliographystyle{IEEEtran}
%\bibliography{lit}

%%%%%%%%%%%%%%%%%%%%%%%%%%%%%%%%%%%%%%%%%%%%%%%%%%%%%%%%%%%%%%%%%%%%%%%%%%%%%%%%%%

%----------------------------------------------------------------------
\begin{IEEEbiographynophoto}{Christian Koller} 
(S'08) received the B.S. and Dipl.-Ing. degrees in electrical engineering from
Technische Universit\"{a}t M\"{u}nchen, Munich, Germany in 2003 and 2004, respectively.
He received the M.S. degree in electrical engineering from University of Notre Dame, 
Notre dame, IN, in 2008.

Since 2005 he is a Research Assistant at the University of Notre Dame, Notre dame, IN,
where he is pursuing a Ph.D. in electrical engineering.
His research interest include error-correcting codes,
network coding, wireless communications, and information theory.
\end{IEEEbiographynophoto}

%----------------------------------------------------------------------
\begin{IEEEbiographynophoto}{Alexandre Graell i Amat} 
(S'01-M'05-SM'10) was born in Barcelona, Catalonia, Spain, on January 28, 1976. 
He received the M.Sc. degree in Telecommunications Engineering from the
Universitat Polit\`{e}cnica de Catalunya, Barcelona, Catalonia,
Spain, in 2001, and the M.Sc. and the Ph.D. degrees in Electrical
Engineering from the Politecnico di Torino, Turin, Italy, in 2000
and 2004, respectively. 

From September 2001 to April 2002, he was a
Visiting Scholar at the Center for Magnetic Recording Research,
University of California at San Diego, La Jolla, CA. From September
2002 to May 2003, he held a visiting appointment at Universitat
Pompeu Fabra, and at the Telecommunications Technological Center of
Catalonia, both in Barcelona. During 2001--2004, he also held a
part-time appointment at STMicroelectronics Data Storage Division,
Milan, Italy, as consultant on coding for magnetic recording
channels. From March 2004 to December 2005, he was a Visiting
Professor at Universitat Pompeu Fabra. 
From January 2006 to December 2010 he was with the Department of 
Electronics of TELECOM Bretagne (former ENST Bretagne), Brest, France. 
In January 2011 he joined the Department of Signals and Systems at 
Chalmers University of Technology, Gothenburg, Sweden, where he is 
currently an Assistant Professor. His research interests are in the 
areas of Coding and Information Theory.

Dr. Graell i Amat is currently an Editor of the IEEE TRANSACTIONS ON COMMUNICATIONS 
and the IEEE COMMUNICATIONS LETTERS. He is the General Co-Chair of the 7th 
International Symposium on Turbo Codes \& Iterative Information Processing, 
Gothenburg, Sweden, August 2012. He received the post-doctoral Juan de la 
Cierva Fellowship of the Spanish Ministry of Education and Science, and the 
Marie Curie Intra-European Fellowship of the European Commission. He was
the runner-up recipient of the IEEE Communications Society ``2010 Europe, 
Middle East \& Africa Region Outstanding Young Researcher Award''.
\end{IEEEbiographynophoto}

%----------------------------------------------------------------------
\begin{IEEEbiographynophoto}{J{\"o}rg Kliewer} 
(S'97--M'99--SM'04) received the Dipl.-Ing.
(M.Sc.) degree in electrical engineering from Hamburg University of
Technology, Hamburg, Germany, in 1993 and the Dr.-Ing. degree
(Ph.D.) in electrical engineering from the University of Kiel, Kiel,
Germany, in 1999, respectively.

From 1993 to 1998, he was a Research Assistant at the University of
Kiel, and from 1999 to 2004, he was a Senior Researcher and Lecturer
with the same institution. In 2004, he visited the University of
Southampton, Southampton, U.K., for one year, and from 2005 until
2007, he was with the University of Notre Dame, Notre Dame, IN, as a
Visiting Assistant Professor. In August 2007, he joined New Mexico
State University, Las Cruces, NM, as an Assistant Professor. His research
interests include network coding, error-correcting codes,
wireless communications, and communication networks.

Dr.~Kliewer was the recipient of a Leverhulme Trust Award and a German
Research Foundation Fellowship Award in 2003 and 2004, respectively.
He was a Member of the Editorial Board of the EURASIP Journal on
Advances in Signal Processing from 2005-2009 and is Associate Editor of the IEEE
Transactions on Communications since 2008.
\end{IEEEbiographynophoto}

%----------------------------------------------------------------------
\begin{IEEEbiographynophoto}{Francesca Vatta} 
received a Laurea in Ingegneria Elettronica in 1992 from University of Trieste, Italy. 
From 1993 to 1994 she has been with Iachello
S.p.A., Olivetti group, Milano, Italy, as system engineer working on design and 
implementation of Computer Integrated Building (CIB) architectures. Since 1995 
she has been with the Department of Electrical Engineering (DEEI) of the University 
of Trieste where she received her Ph.D. degree in Telecommunications in 1998, with 
a Ph.D. thesis concerning the study and design of source-matched channel coding 
schemes for mobile communications. In November 1999 she became assistant professor 
at the University of Trieste. Starting in 2002, she spent several months as visiting 
scholar at the University of Notre Dame, Notre Dame, IN, U.S.A., cooperating with 
the Coding Theory Research Group under the guidance of Prof. D. J. Costello, Jr. 
Starting in 2005, she spent several months as visiting scholar at the University of 
Ulm, Germany, cooperating with the Telecommunications and Applied Information Theory 
Research Group under the guidance of Prof. M. Bossert. She is an author of more than 
70 papers published on international journals and conference proceedings. Her current 
research interests are in the area of channel coding concerning, in particular, the 
analysis and design of concatenated coding schemes for wireless applications.
\end{IEEEbiographynophoto}

%----------------------------------------------------------------------
\begin{IEEEbiographynophoto}{Kamil Sh. Zigangirov} 
(M'95–-SM'99–-F'01) was born in the U.S.S.R. in 1938. He
received the M.S. degree in 1962 from the Moscow Institute for Physics and
Technology,Moscow, U.S.S.R., and the Ph.D. degree in 1966 from the Institute
of Radio Engineering and Electronics of the U.S.S.R. Academy of Sciences,
Moscow.

From 1965 to 1991, he held various research positions with the Institute for
Problems of Information Transmission of the U.S.S.R. Academy of Sciences,
Moscow, first as a Junior Scientist, and later as a Main Scientist. During this
period, he visited several universities in the United States, Sweden, Italy, and
Switzerland as a Guest Researcher. He organized several symposia on information
theory in the U.S.S.R. In 1994, he received the Chair of Telecommunication
Theory at Lund University, Lund, Sweden. From 2003 to 2009, he was a Visiting
Professor with the University of Notre Dame, Notre Dame, IN, the Dresden
Technical University, Dresden, Germany, and with the University of Alberta,
Edmonton, Alberta, Canada. His scientific interests include information theory,
coding theory, detection theory, and mathematical statistics. In addition to papers
in these areas, he published a book on sequential decoding of convolutional
codes (in Russian) in 1974. With R. Johannesson, he coauthored the textbook
\emph{Fundamentals of Convolutional Coding} (IEEE, 1999) and authored \emph{Theory of
CDMA Communication} (IEEE, 2004).
\end{IEEEbiographynophoto}

%----------------------------------------------------------------------
\begin{IEEEbiographynophoto}{Daniel J. Costello, Jr.} 
(S'62–-M'69–-SM'78–-F'86–-LF'08) was born in Seattle, WA, on August 9, 1942. He 
received the B.S.E.E. degree from Seattle University,
Seattle, WA, in 1964, and the M.S. and Ph.D. degrees in electrical engineering
from the University of Notre Dame, Notre Dame, IN, in 1966 and 1969,
respectively.

He joined the faculty of the Illinois Institute of Technology, Chicago, in
1969 as an Assistant Professor of Electrical Engineering. He was promoted
to Associate Professor in 1973, and to Full Professor in 1980. In 1985 he
became Professor of Electrical Engineering with the University of Notre
Dame, and from 1989 to 1998, served as Chair of the Department of Electrical
Engineering. His research interests are in the area of digital communications,
with special emphasis on error control coding and coded modulation. He has
numerous technical publications in his field, and in 1983 he coauthored a
textbook entitled \emph{Error Control Coding: Fundamentals and Applications}, the
2nd edition of which was published in 2004.

Dr. Costello has been a member of the Information Theory Society Board of
Governors (BOG) since 1983, and in 1986, he served as President of the BOG.
He has also served as Associate Editor for Communication Theory for the IEEE
TRANSACTIONS ON COMMUNICATIONS, Associate Editor for Coding Techniques
for the IEEE TRANSACTIONS ON INFORMATION THEORY, and Co-Chair of the
IEEE International Symposia on Information Theory in Kobe, Japan (1988),
Ulm, Germany (1997), and Chicago, IL (2004). In 1991, he was selected as one
of 100 Seattle University alumni to receive the Centennial Alumni Award in
recognition of alumni who have displayed outstanding service to others, exceptional
leadership, or uncommon achievement. In 1999, he received a Humboldt
Research Prize from the Alexander von Humboldt Foundation in Germany. In
2000, he was named the Leonard Bettex Professor of Electrical Engineering at
the University of Notre Dame. In 2000, he was selected by the IEEE Information
Theory Society as a recipient of a Third-Millennium Medal. He was corecipient
of the 2009 IEEE Donald G. Fink Prize Paper Award, which recognizes an outstanding
survey, review, or tutorial paper in any IEEE publication issued during
the previous calendar year.
\end{IEEEbiographynophoto}
%-----------------------------------------------------------------------------------------

\end{document}